\documentclass{nature}
\usepackage{graphicx}
\usepackage{amsmath}
\usepackage{amsxtra}
\usepackage{amstext}
\usepackage{amssymb}
\usepackage{caption}
\usepackage{lineno}
\usepackage{upgreek}
\usepackage{color}

\title{Multimode strong coupling in cavity optomechanics} 

\author{Prashanta Kharel$^{1*}$, Yiwen Chu$^{1*}$, Eric A. Kittlaus$^{1}$, Nils T. Otterstrom$^1$, Shai Gertler$^1$, \& Peter T. Rakich$^{1}$}

\begin{document}

\maketitle

\begin{affiliations}
\item Department of Applied Physics, Yale University, New  Haven, Connecticut 06511, USA and
Yale  Quantum  Institute,  Yale  University,  New  Haven,  Connecticut  06520,  USA
\end{affiliations}

\begin{abstract}

The field of optomechanics has successfully harnessed coupling between light and mechanical motion to perform precision measurements\cite{schnabel2010quantum, schreppler2014optically}, create complex quantum states\cite{riedinger2018remote, wollman2015quantum}, and develop new technologies\cite{yu2016cavity}. Building on these accomplishments, optomechanical systems show great potential as quantum transducers and information storage devices for use in future hybrid quantum networks\cite{kimble2008quantum} and offer novel strategies for quantum state preparation to explore macroscopic quantum phenomena\cite{clarke2018growing}. Towards these goals, deterministic control of optomechanical interactions in the strong coupling regime\cite{aspelmeyer2014cavity} represents an important strategy for efficient utilization of quantum degrees of freedom in mechanical systems. While strong coupling has been demonstrated in both electromechanical\cite{o2010quantum, teufel2011circuit} and optomechanical\cite{groblacher2009observation, verhagen2012quantum,enzian2019observation} systems, it has proven difficult to identify a robust optomechanical system that features the low loss and high coupling rates required for more sophisticated control of mechanical motion. In this paper, we demonstrate robust strong coupling between multiple long-lived phonon modes of a bulk acoustic wave (BAW) resonator and a single optical cavity mode. We show that this so-called ``multimode strong coupling" regime can be a powerful tool to shape and control decoherence pathways through nontrivial forms of mode hybridization. Using a combination of frequency- and time-domain measurements, we identify hybridized modes with lifetimes that are significantly longer than that of any mode of the uncoupled system. This surprising effect, which results from the interference of decay channels, showcases the use of multimode strong coupling as a general strategy to mitigate extrinsic loss mechanisms. Moreover, the phonons supported by BAW resonators have a collection of properties, including high frequencies\cite{renninger2018bulk}, long coherence times\cite{galliou2013extremely}, and robustness against thermal decoherence\cite{chu2018creation}, making this optomechanical system particularly enticing for applications such as quantum transduction and memories\cite{bochmann2013nanomechanical,andrews2014bidirectional}. These results show that our system can be used to study novel phenomena in a previously unexplored regime of optomechanics and could be an important building block for future quantum devices.

\end{abstract}

As is the case for many quantum-optical systems, optomechanical devices exhibit novel physical behaviors and acquire new useful capabilities when they enter the so-called ``strong coupling regime"\cite{aspelmeyer2014cavity}. In this regime, the coupling rate between light and motion becomes faster than both the optical and mechanical dissipation rates, which is necessary for applications such as quantum memories and transduction. Since light is the natural carrier of quantum information over long distances, and mechanical motion efficiently couples to many quantum systems, a robust and coherent interface between light and mechanical motion could be a useful building block in hybrid quantum networks for long-distance communications\cite{stannigel2010optomechanical} or modular quantum computation\cite{lee2012macroscopic}.

Only a few optomechanical devices have entered the regime of optomechanical strong coupling due to technical challenges associated with realizing low-loss systems that can also robustly support high coupling rates. Radiation pressure has been used to achieve strong coupling between THz-frequency optical modes and MHz-frequency mechanical modes within micromechanical systems\cite{groblacher2009observation,verhagen2012quantum}. However, if we seek to utilize optomechanical systems as a quantum resource, it is advantageous to instead utilize high frequency (GHz) phonon modes; this is because higher frequencies improve robustness to thermal decoherence, enable faster quantum operations, and permit ground state operation at cryogenic temperatures. Despite the many successes of GHz-frequency micro- and nano-optomechanical systems\cite{eichenfield2009optomechanical}, it remains challenging to reach strong coupling in such systems due to practical limits\cite{chan2011laser,riedinger2016non} on the circulating photon number, which in turn limit the cavity-enhanced coupling rate.

Through an alternative approach, Brillouin interactions have been used to demonstrate strong coupling to GHz-frequency mechanical modes of a fused-silica whispering gallery mode resonator\cite{enzian2019observation}. This strategy permits resonant driving of the optical mode, and the macroscopic size of the system alleviates some of the technical challenges associated with laser heating, making large coupling rates more readily accessible. While this recent demonstration illustrates the advantages of Brillouin-based coupling in macroscopic systems, low acoustic dissipation rates--necessary to store information in the mechanical mode--are difficult to achieve in glasses at cryogenic temperatures. In particular, two-level tunnelling-state systems, which are intrinsic to silica, produce excess dissipation and noise at cryogenic temperatures\cite{hunklinger1976ultrasonic,arcizet2009cryogenic}, complicating the prospects for efficient quantum operations in such systems. However, a promising way to address this challenge could be to realize strong coupling between optical cavity modes and the long-lived high-frequency ($>$10 GHz) phonon modes supported by crystalline bulk acoustic wave (BAW) resonators\cite{kharel2018high}.

Separate from the opportunities presented by strong coupling, another direction with significant untapped potential in cavity optomechanics--more generally in quantum information science--involves the exploration of coupled multimode systems. Such systems have already given rise to the observation of a wide variety of interesting physical phenomena, such as optomechanical dark modes\cite{lin2010coherent,dong2012optomechanical}, synchronization of mechanical frequencies\cite{heinrich2011collective}, topological dynamics\cite{xu2016topological}, and non-reciprocity\cite{ruesink2016nonreciprocity}. Achieving strong coupling within a multimode system could lead to a wealth of new capabilities including storing light through control of bright and dark states\cite{massel2012multimode} and exploration of topological phonon transport\cite{peano2015topological}. Furthermore, in the quantum regime, multimode strong coupling could open the door to generation of multipartite mechanical entanglement\cite{weaver2018phonon} and the implementation of a quantum simulator for many-body bosonic systems\cite{ludwig2013quantum, Hartmann2016Quantum}.

In this Letter, we demonstrate multimode strong coupling in a high-frequency (12.6 GHz) cavity optomechanical system at cryogenic temperatures. Our system combines a high-finesse optical resonator with a low-loss, crystalline BAW resonator, which can be reconfigured so that the optical mode is strongly coupled to either a single acoustic mode or several acoustic modes. Using both frequency and time-domain measurements, we quantify the parameters of the system and explore its dynamics. In the single mode case, we achieve an optomechanical coupling rate of $g_{\rm m} = 2 \pi \times (7.2 \pm 0.1)$ MHz, which exceeds both the optical dissipation rate of $\kappa = 2 \pi \times (4.43\pm0.02)$ MHz and the mechanical dissipation rate of $\Gamma_{\rm m } = 2 \pi \times  (66\pm 3)$ kHz.
In the multimode case, the coupling rates exceed the acoustic free spectral range of $\delta = 2\pi \times (610\pm 10)$ kHz, meaning we enter the multimode strong coupling regime. 
In this regime, we observe the formation of optomechanical ``dark modes" with linewidths that are a factor of five less than the smallest dissipation rate, $\Gamma_{\rm m}$, of the uncoupled system. We show that this intriguing phenomenon can be explained by the destructive interference of radiative loss channels for the dark modes.

Our optomechanical system consists of a planar quartz crystal that is placed within a Fabry-P\'erot optical cavity with high reflectivity (99.9\%) optical mirrors (Fig. \ref{fig1}a). At a temperature of $\sim$ 10 K, long-lived longitudinal acoustic modes within the quartz crystal are reflected from the planar surfaces of the crystal to form a series of macroscopic standing wave acoustic modes similar to the standing wave electromagnetic modes formed within a Fabry-P\'erot optical cavity. A high-frequency acoustic mode within the BAW resonator (formed by the quartz crystal) can mediate coupling between two distinct longitudinal modes of the optical cavity through Brillouin interactions when energy conservation and phase matching requirements are satisfied (Fig. \ref{fig2}b).
For crystalline $z$-cut quartz at cryogenic temperatures and optical modes near 1550 nm, such interactions occur for a narrow band of acoustic modes near 12.6 GHz (see Supplementary Information section 1). 

This multimode coupling can be described by the interaction Hamiltonian
\begin{align}
H_\text{int} = -\sum_\text{m}\hbar g_0^\text{m}(a_2^{\dagger} a_1 b_\text{m} + a_1^{\dagger} a_2 b_\text{m}^{\dagger}),
\end{align}
where $a_1^\dagger$ ($a_2^\dagger$) is the creation operator for the optical mode at frequency $\omega_1$ ($\omega_2$), $b_\text{m}^{\dagger}$ is the creation operator for the acoustic mode at frequency $\Omega_\text{m},$ and $g_0^\text{m}$ is the zero-point coupling rate. We  note that $g_0^\text{m}$ depends on the spatial acousto-optical overlap, which provides us with a way of tailoring the optomechanical coupling strength for different acoustic modes\cite{kharel2018high}. With an external control laser that is driven on resonance with the lower frequency optical mode $a_1$, we can write an effective linearized Hamiltonian as 
\begin{align} \label{Heff} 
H_\text{eff} = \hbar \Delta a_2^{\dagger}a_2+\sum_\text{m}\hbar \Omega_\text{m}b_\text{m}^{\dagger} b_\text{m}  -   \sum_\text{m}\hbar g_\text{m} (a_2^{\dagger} b_\text{m} + a_2 b_\text{m}^{\dagger}).
\end{align}
Here, we have moved to the rotating frame of mode $a_1$, $g_m= \sqrt{\bar{n}_\text{c}} g_0^\text{m}$ is the cavity-enhanced coupling rate, $\bar{n}_c$ is the intra-cavity photon number of mode $a_1$, and $\Delta = \omega_2-\omega_1$ is the optical free spectral range. 

The above beam-splitter Hamiltonian $\hbar g_\text{m} (a_2^{\dagger} b_\text{m} + a_2 b_\text{m}^{\dagger})$ describes coherent energy exchange between a single optical mode $a_2$ and a single acoustic mode $b_\text{m}$ with interaction rate $ 2 g_\text{m}$. However, the dissipation rates relative to this interaction rate determine the optical transmission spectrum of mode $a_2$, which we measure using a weak probe field. In the weak coupling regime $g_\text{m} < ( \kappa/2, \Gamma_\text{m}/2)$, 
we expect to observe a narrow dip in the transmission spectrum due to the well known phenomenon of optomechanically induced transparency (OMIT)\cite{weis2010optomechanically} seen in Fig. \ref{fig1}c. In the strong coupling regime $g_\text{m} > ( \kappa/2, \Gamma_\text{m}/2),$ the optical transmission spectrum develops two resonant features that correspond to new modes that result from the hybridization between the optical mode $a_2$ and the individual mechanical mode $b_{\rm m}$ seen in Fig. \ref{fig1}c.


Because our experimental system permits coupling to an array of acoustic modes, the optical mode spectrum develops additional features in the strong coupling regime. Since the BAW resonator supports multiple acoustic modes with regular frequency spacing ($\delta$) that is smaller than the optical dissipation rate ($\kappa$), it is important to go beyond the minimal model of a single optical mode coupled to a single phonon mode. This is because, more than one acoustic modes can simultaneously mediate coupling between the same pair of optical modes (Fig. \ref{fig1}b). Therefore, in addition to normal-mode splitting of a single strongly coupled acoustic mode, we expect several OMIT dips to arise from weak coupling to a multitude of acoustic modes as seen in Fig. \ref{fig1}d.

We first present experimental measurements of strong optomechanical coupling when our system is configured to couple predominantly to a single acoustic mode. For the lowest control laser power, the transmission spectrum seen in Fig. \ref{fig2}a.i  reveals a single OMIT dip at $\Omega_\text{m} = 2 \pi \times 12.591$ GHz. Through these low power measurements in the weak coupling regime, we extract $\kappa = 2\pi \times( 4.43 \pm 0.02)$ MHz, $\Gamma_\text{m} = 2\pi \times ( 66 \pm 3)$ kHz, and $g_0^m= 2\pi \times ( 23 \pm 1)$ Hz (See Supplementary Information section 2). To reach the regime of strong coupling, we enhance $g_\text{m}$ by increasing $\bar{n}_\text{c}$. As expected from theory, we observe a normal-mode splitting that increases proportional to $\sqrt{P_\text{in}}$ (Fig. \ref{fig2}a), where $P_\text{in}$ is the input control laser power. At the highest $P_\text{in}$ of 187 mW, corresponding to an intracavity photon number $\bar{n}_\text{c}= 1.1 \times 10^{11}$, we observe a splitting $2 g_{\rm m}= 2\pi \times (14.4 \pm 0.1)$ MHz (Fig. \ref{fig2}a.ii). Since $2g_\text{m}/\kappa \simeq 3$ and $2g_\text{m}/\Gamma_\text{m} \simeq 220 $, the coherent coupling rate far exceeds the dissipation rates of both the optical and the acoustic modes, indicating that our system is in the strong coupling regime.
Given that the thermal occupation for the phonon mode $\bar{n}_\text{th} \simeq 25$, we find that the coherent coupling rate, $2g_{\rm m}$, is more than eight times larger than the mechanical thermal decoherence rate, $\gamma_{\rm m} = \bar{n}_\text{th} \Gamma_\text{m}$, establishing that our system is also in the quantum-coherent strong-coupling regime\cite{verhagen2012quantum} (see Supplementary Information section 3). 


As another unambiguous signature of strong coupling, we tune the FSR of the optical cavity modes into and out of resonance with the strongly coupled acoustic mode to reveal a characteristic anticrossing feature (Fig. \ref{fig2}b). Because the optical mode spacing, $\Delta$, depends on the temperature, $T$, we can readily tune $\Delta$ to match the frequency, $\Omega_{\rm m}$, of Brillouin-active phonon modes. During these measurements, we lock the control laser on resonance with the optical mode at $\omega_1$, such that temperature tuning allows us to change optical mode spacing, $\Delta$, without changing $g_\text{m}$. From transmission spectra obtained as a function of $T$ at the highest control power, we observe a clear anticrossing at $T=7.6$ K when $\Delta \simeq \Omega_\text{m}$ (Fig. \ref{fig2}b.i). For the off-resonant case ($\Delta \neq \Omega_\text{m}$) at $T=12.4$ K, we obtain a narrow and a broad resonance features seen in Fig. \ref{fig2}b.ii, which correspond to the acoustic and the optical modes of the uncoupled system, respectively.

To complement the frequency-domain measurements described above, we also demonstrate time-domain control of the system and identify salient features of strong coupling in the measured response. Through time domain-measurements, we pulse the weak probe light ($\omega_2$) while maintaining a strong continuous drive at $\omega_1$ (Fig. \ref{fig3}a). A heterodyne signal resulting from the interference between the control and the probe light transmitted through the cavity provides phase-sensitive detection of the probe light as a function of time (See Supplementary Information section 1). These time-domain measurements, shown in Fig. \ref{fig3}b, were performed at the same temperatures as the frequency-domain measurements shown in Fig. \ref{fig2}b, and result in a characteristic detuning dependency of Rabi oscillations obtained when two resonators coherently exchange energy in the strong coupling regime\cite{chu2017quantum}. At $T =$ 7.6 K (when $\Delta \simeq \Omega_\text{m}$), we observe coherent oscillations with a period of 69 ns, which is consistent with the value of $2 g_\text{m}$ extracted from frequency domain measurements. As seen in Fig. \ref{fig3}c the time constant $\tau$ for this energy decay is 70 ns, which agrees well with the energy decay rate of $(\kappa+\Gamma_\text{m})/2 \approx \kappa/2$ of the hybridized modes. Notice, however, that additional revivals of the coherent oscillations are observed for $t \gg \tau$ in Fig. \ref{fig3}c. These nontrivial features appear because the spectrally-broad probe pulse excites a single strongly coupled acoustic mode as well as a multitude of weakly coupled acoustic modes that lie outside the phase matching bandwidth. Due to the modulation in the coupling strength produced by the phase matching conditions, $g_0^m$ is suppressed for alternating acoustic mode numbers $m$ outside the phase matching bandwidth (Fig. \ref{fig1}d). For this reason, the observed revivals have a period of $0.82\  \upmu $s, corresponding to a frequency of $1.2$ MHz, which is approximately twice the acoustic free-spectral range of 610 kHz. Since the lifetimes of such weakly coupled modes approach the intrinsic mechanical decay time of the uncoupled system, $1/\Gamma_\text{m} \simeq 2.4 \ \upmu$s, the revivals are sustained for $t \gg \tau$.

To further explore such multimode dynamics in this optomechanical system, we configure our system so that one optical mode strongly couples to three acoustic modes (Fig. \ref{fig4}a). This is accomplished by tuning the optical wavelength to select a different pair of optical modes, which changes the spatial overlap between the optical and acoustic modes (See Supplementary Information section 1). The transmission spectrum taken at low power (Fig. \ref{fig4}b.i) reveals three OMIT dips. As before, theoretical fits to OMIT spectrum at low powers allow us to extract coupling rates $g_1= 2\pi \times (4.9 \pm 0.1)$ MHz, $g_2= 2\pi \times (4.0 \pm 0.1)$ MHz and $g_3= 2\pi \times (3.7 \pm 0.1)$ MHz, as well as dissipation rates $\kappa = 2\pi \times  (2.52 \pm 0.08)$ MHz and $\Gamma_\text{m} = 2\pi \ \times (67 \pm 10)$ kHz (See Supplementary Information section 2). In the strong coupling regime, we observe four distinct peaks in the transmission spectrum seen in Fig. \ref{fig4}b.ii. These peaks represent the four eigenmodes produced by the hybridization of the optical mode ($a_2$) with the three dominant phonon modes $b_1$, $b_2$, and $b_3$. 

To understand the nature of these four new eigenmodes, we start by considering a simpler case of a single optical mode ($a_2$) coupled strongly to two phonon modes ($b_1$, $b_2$) separated by $2\delta$. Furthermore, we assume that $g_1 = g_2 \equiv g$, $\Gamma_1= \Gamma_2 \equiv\Gamma$, and $\Omega_{1,2} = \Delta \mp \delta$. The Hamiltonian of this three coupled oscillator system in the basis of $a_2$, $b_1$, and $b_2$ is given by
\begin{equation} \label{Heff2phonon}
H_\text{eff}=
  \left[\begin{array}{ccc}    
    \Delta - i \kappa/2 & -g & -g \\
    -g^* & \Omega_1-i\Gamma/2 & 0 \\
    -g^* & 0 & \Omega_2-i\Gamma/2 
    \end{array}\right].
\end{equation}
This effective Hamiltonian can be diagonalized to obtain three eigenmodes of the hybridized system (see Supplementary Information section 5). In the limit of large $g$, these eigenmodes become two `bright' modes $B_{\pm} = \frac{1}{\sqrt{2}} a_2 \pm \frac{1}{2} (b_1+b_2)$ at frequencies $\omega_{\pm} = \Delta \pm \sqrt{2} g$ with dissipation rates  $\kappa_{\pm} = \kappa/2$ and one `dark' mode $D = \frac{1}{\sqrt{2}} (b_1-b_2)$ at frequency $\omega_\text{D}=\Delta$ with a dissipation rate $\kappa_\text{D} = \Gamma$. Notice that the bright modes are formed from the superposition of both the optical and the acoustic modes whereas the dark mode lacks an optical mode component, meaning that it does not couple to light. The dynamics of such a system, and the existence of such bright and dark modes, has been explored in an electromechanical system using a GHz frequency microwave resonator strongly coupled to two MHz frequency micromechanical oscillators\cite{massel2012multimode}. However, this regime of coupling has not been previously accessible for optomechanical systems. 


From a straightforward generalization of the effective Hamiltonian in Eq. (\ref{Heff2phonon}) to the case of 3 phonon modes, we now expect four eigenmodes of the hybridized system seen in Fig. \ref{fig4}a. Of these eigenmodes, the two broad peaks correspond to the bright modes, whereas the two narrow peaks correspond to the dark modes. We expect the decay rates of these dark modes to approach the mechanical decay rate $\Gamma_{\rm m} = 2\pi \times $ 67 kHz. However, high-resolution measurements of such modes at the highest control laser powers (seen in Fig. \ref{fig4}b.iv) reveal decay rates $\Gamma_\text{d2}= 2\pi \ \times 14$ kHz and $\Gamma_\text{d3}= 2\pi \ \times 15$ kHz, which are approximately 5 times smaller than the original acoustic dissipation rate $\Gamma_{\rm m}$. Time-domain ring-down measurements using probe pulses with a narrow spectral bandwidth centered around each dark mode confirmed their long-lived nature (Fig. \ref{fig4}c). The measured decay time of  $\tau_d \sim 10.9 \ \upmu$s for both modes is consistent with the linewidths. This linewidth-narrowing phenomena is surprising because these eigenmodes, which are hybridized excitations of light and sound, have decay rates that are smaller than the optical and the mechanical decay rates of the uncoupled system.

The observed line-narrowing phenomenon for the dark modes in our optomechanical system can be understood as interference in decay pathways resulting from the decay of mechanical modes into a common reservoir (see Supplementary Information section 5). The acoustic loss in our system is dominated by radiative loss mechanisms, which include diffraction and beam walk-off\cite{kharel2018high}. Therefore, longitudinal acoustic modes, with nearly identical Gaussian transverse mode profiles (determined by the optical field), decay into a common continuum of higher-order transverse modes that span the entire crystal. Since the dark mode is an anti-symmetric superposition of two acoustic modes, the decay pathways can destructively interfere and lead to line-narrowing beyond that associated with radiative loss mechanisms. 

This interference phenomenon can be described by including additional dissipative coupling terms\cite{metelmann2015nonreciprocal, fang2017generalized} in the effective Hamiltonian of the form $\text{H}_{\text{nm}} = -i(\Gamma_{\text{nm}}/2)b_\text{n}^{\dagger}b_\text{m}$, where $\Gamma_{\rm nm}$ results only from the radiative part of the dissipation. In the simple case of two acoustic modes treated above, the resulting decay rate of the dark mode, $D$, becomes $\kappa_\text{D} = (\Gamma-\Gamma_{12})+ \kappa \delta^2/2g^2$. Therefore, we expect the linewidth of the dark mode to decrease with a $ 1/g^2$ proportionality, approaching $(\Gamma-\Gamma_{12})/2\pi$ as $g$ becomes large. Additionally, if the acoustic dissipation of the two modes were equal, and entirely due to radiative loss, one finds $\Gamma_{12}  = \Gamma$ and $\kappa_\text{D} \rightarrow 0$. However, in addition to geometric loss, acoustic modes in our system also suffer from intrinsic dissipation mechanisms such as scattering and absorption due to imperfections in the crystal (i.e., $\Gamma_i \equiv \Gamma -\Gamma_{12} \neq 0$). Hence, for large $g$, we expect the dissipation rates of the dark modes to approach the intrinsic decay rate $\Gamma_i$, which can be very small for acoustic waves in pristine crystalline media at cryogenic temperatures\cite{renninger2018bulk,galliou2013extremely}. We note that analogous line-narrowing phenomena due to interference of decay pathways has been investigated in fluorescence spectra of a V-type atomic system\cite{zhou1997quantum} as well as in a circuit quantum electrodynamics (cQED) platform\cite{sundaresan2015beyond}. 

Measurement of the linewidths of the two dark modes as a function of power (Fig. \ref{fig4}d) agrees well with the theoretical description of our system presented above. Theoretical fits to the data, for the case of three phonon modes, were performed by numerically diagonalizing the effective Hamiltonian that includes the dissipative coupling terms $\Gamma_{\text{nm}}$. Only $\Gamma_i= 2\pi \times 5 $ kHz is taken as a fit parameter (see Supplementary Information section 5) and is consistent with independent measurements of acoustic damping in quartz crystals at cryogenic temperatures\cite{renninger2018bulk, kharel2018ultra}. Note that we observe a larger than expected linewidth of the dark modes at the highest powers, which could be due to a deviation from the linear dependence of $g$ on $\sqrt{P_\text{in}}$ (see Supplementary Information section 3).

These results demonstrate that optomechanical coupling to a multitude of high-frequency, low-loss phonons within BAW resonators present intriguing new opportunities for novel classical and quantum phenomena in the multimode strong coupling regime. Moreover, the time-domain measurements presented in this paper represent an important step towards optical control of bulk acoustic phonons for quantum transduction and the generation of non-classical mechanical states. While our system is already in the quantum-coherent strong-coupling regime necessary to observe quantum effects, a number of improvements can be made to achieve robust quantum control of phonons and realize the aforementioned goals. First, it is possible to directly initialize such high-frequency (12.6 GHz) phonons in their quantum ground states at temperatures $<1$ K by using a standard dilution refrigerator. Decreasing $\kappa$ by improving mirror reflectivity to 99.99 $\%$ and utilizing low-loss crystalline substrates with larger Brillouin gain (such as TeO$_2$) could enable access to the strong coupling regime at $<100 \ \upmu$W input powers. These improvements, along with low duty-cycle pulsed operation of the control laser with micro-Watt average powers could make operation in dilution refrigerators feasible.  More importantly, it would be beneficial to design a fiber-coupled optical cavity system with piezo-tunable crystal position to minimize stray optical reflections as well as to provide enhanced control over coupling to one or more phonon modes.

These improvements could offer new avenues for utilizing multimode optomechanical interactions for future applications in quantum information and metrology. For instance, it may be possible to adiabatically transfer quantum optical states, such as single-photons, to the long-lived dark states for quantum information storage. Moreover, it may be possible to generate entangled mechanical states by swapping a single optical excitation simultaneously to two strongly coupled acoustic modes. In addition, as shown by the line-narrowing phenomena we observed, strong coupling between light and acoustic modes within BAW resonators could be used to explore and mitigate acoustic dissipation mechanisms. More generally, it has been shown that acoustic waves within BAW resonators couple strongly to a variety of other quantum systems such as superconducting qubits\cite{chu2018creation}, defect centers\cite{macquarrie2015coherent} and microwave fields\cite{han2016multimode}. Therefore, deterministic control of bulk acoustic waves using light in the strong coupling regime could be a valuable tool for manipulating quantum information and exploring new physical phenomena in hybrid quantum systems.


\begin{addendum}
 \item The authors thank Jack G. E. Harris and Robert J. Schoelkopf for insightful discussions regarding coherent phenomena and for generous contribution of technical expertise and experimental resources. 
 We also thank Luke Burkhart, Vijay Jain, and Yizhi Luo for helpful discussions and valuable feedback.
 We acknowledge funding  support  from  ONR  YIP (N00014-17-1-2514), the US Army Research Office (W911NF-14-1-0011), and the Packard Fellowship for Science and Engineering. N.T.O. acknowledges support from the National Science Foundation Graduate Research Fellowship under Grant No.  DGE1122492.  The authors of this paper are contributors to patent application no.  62/465101 related to Bulk Crystalline Optomechanics, which was submitted by Yale University.
 
 \item[Author contributions] P.K., Y.C., and E.A.K. performed the experiments under supervision of P.T.R.. P.K. and Y.C. analyzed the data and developed analytical theory under the guidance of P.T.R.. P.K. designed and built the experimental apparatus to perform the measurement with support from Y.C. and P.T.R.. E.A.K., N.T.O., and S.G. aided in the development of experimental techniques. All authors participated in the writing of this manuscript. P.K and Y.C. contributed equally to this work.
 \item[Author information] Correspondence and requests for materials should be addressed to P. Kharel (email: prashanta.kharel@yale.edu) or P. T. Rakich (peter.rakich@yale.edu ).
 \item[Data availability] All data needed to evaluate the conclusions in the paper are present in the paper and/or the supplementary materials. Additional data are available from the corresponding authors upon reasonable request.

\end{addendum}

\newpage
\begin{figure}[]
\centerline{
\includegraphics[width=130 mm]{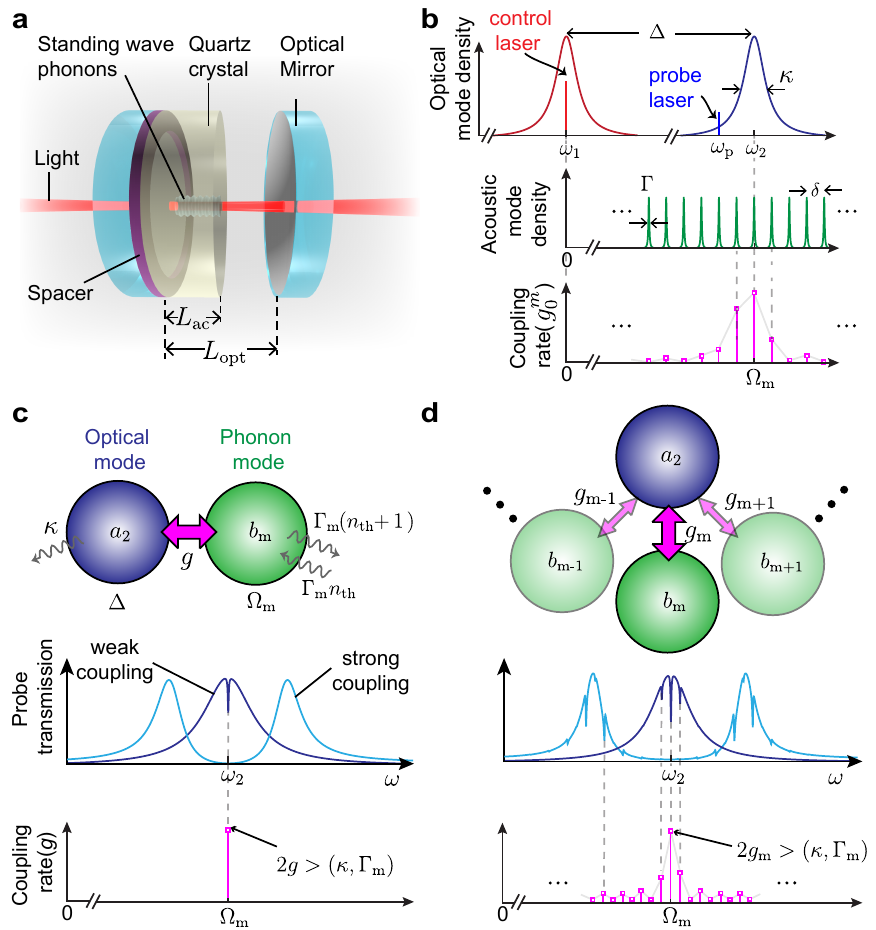}}
\end{figure} 
\captionof{figure}{ \textbf{Multimode cavity optomechanical system.} \textbf{a}, Schematic of the optomechanical system (not to scale). The thickness of the half-inch diameter quartz crystal is $L_{\rm ac} = 5$ mm and the spacing between the optical mirrors is $L_{\rm opt} = $ 9.9 mm. The optical and acoustic mode waist diameter is 122 $\upmu$m and  86 $\upmu$m, respectively.  \textbf{b}, Schematic spectra of optical modes (top) and acoustic modes (middle). The zero-point optomechanical coupling rates (bottom) are determined by a combination of the Brillouin bandwidth dictated by energy and momentum conservation, and the spatial overlap of acoustic and optical modes\cite{kharel2018high}. \textbf{c}, Diagram of linearized optomechanical coupling between an optical mode and a single acoustic mode (top), corresponding to expected spectra of probe laser transmission in the weak- and strong-coupling regime (middle), and coupling rate under a strong control laser drive (bottom). \textbf{d}, Same as in \textbf{c}, except the optical mode is coupled to many acoustic modes. Here, the coupling to one acoustic mode is dominant, corresponding to the case in Figures \ref{fig2} and \ref{fig3}.
}
\label{fig1}

\newpage
\begin{figure}[]
\centerline{
\includegraphics[width=183 mm]{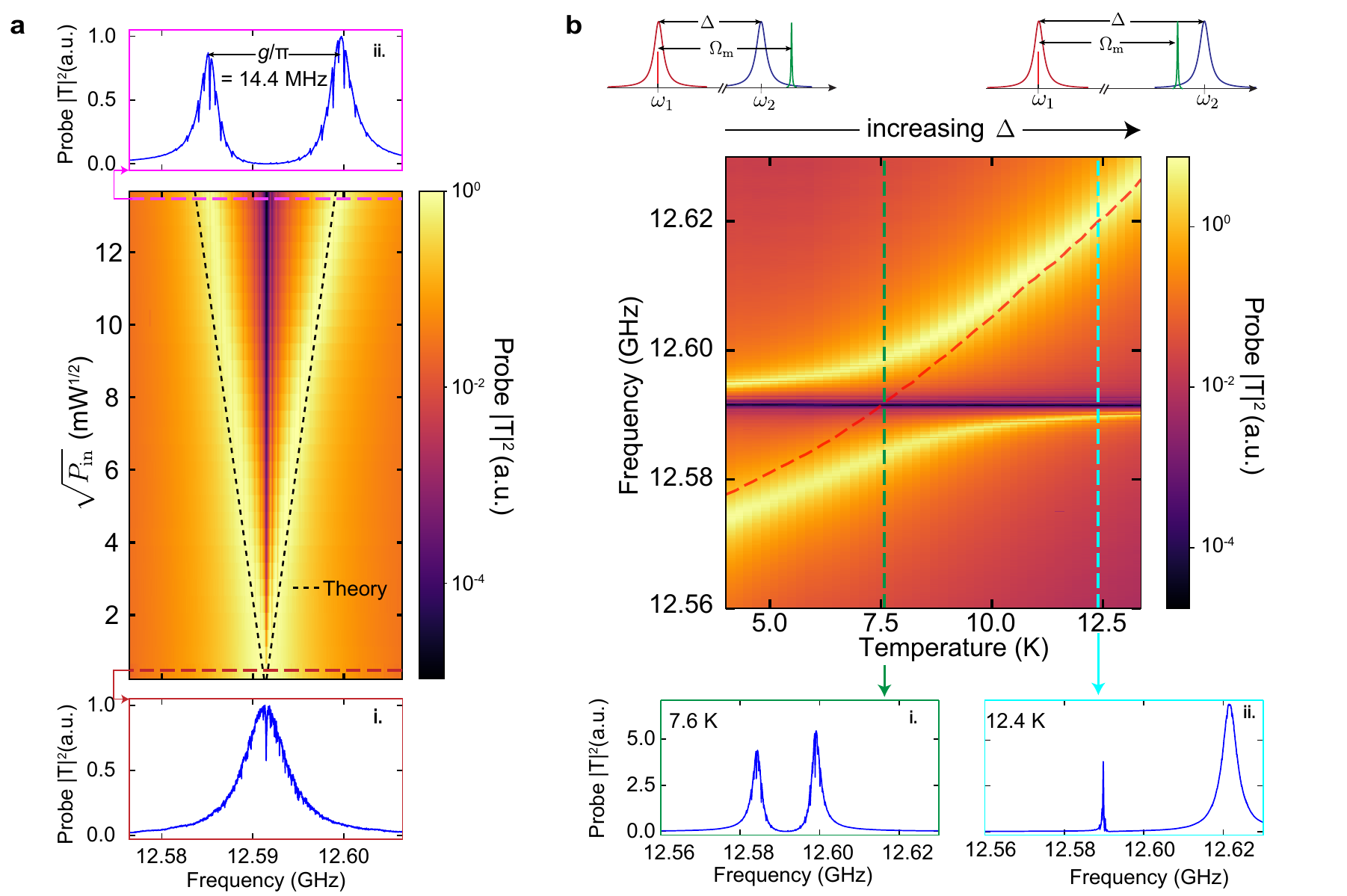}}
\end{figure}  
\captionof{figure}{ \textbf{Optomechanical strong coupling to a single mechanical mode.} \textbf{a}, Probe laser transmission spectra taken at various control laser powers. Dashed lines show expected values of $g_{\rm m}$ for the dominantly coupled acoustic mode, extrapolated from fits to low-power spectra with $\sqrt{P_{\rm in}} < 0.6 \text{ mW}^{1/2}$ (see Supplementary Information section 2). Lower and upper panels show spectra in the regimes of weak and strong coupling, respectively. In the strongly coupled case, the normal-mode splitting indicated is due to the dominantly-coupled acoustic mode, but OMIT features are visible from other acoustic modes that are still weakly coupled. \textbf{b}, Probe transmission taken at various cryostat temperatures. Dashed red line shows the value of $\Delta$ at each temperature, obtained by fitting the data to the theoretical expression for probe transmission (see Supplementary Information section 2). Lower panels show spectra in the resonant and far-detuned cases.
}
\label{fig2}

\begin{figure}[]
\centerline{
\includegraphics[width=130mm]{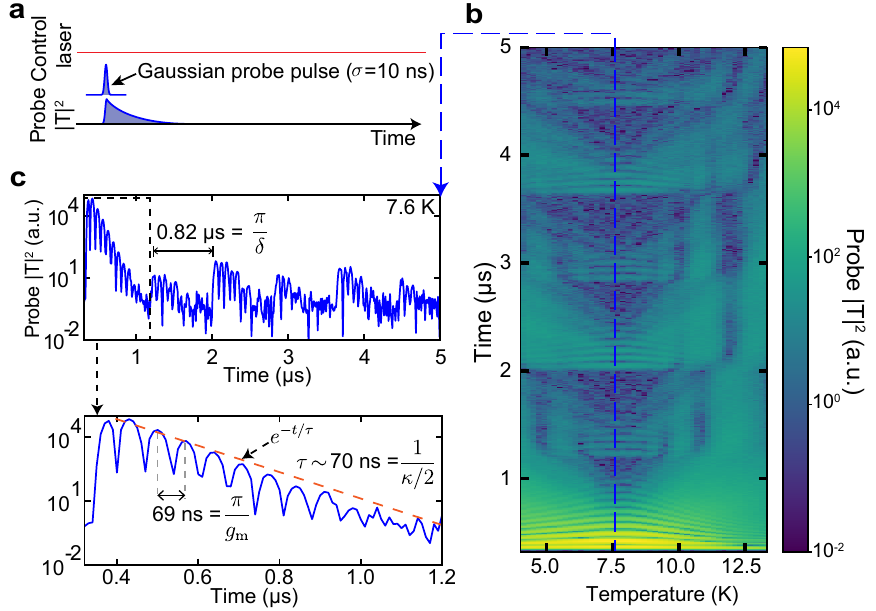}}
\caption{ \textbf{Time-domain measurements of strong coupling.} \textbf{a}, Schematic of the time-domain measurement. A strong control laser is continuously on-resonance with the optical mode at $\omega_1$ to turn on the optomechanical coupling (see Fig. \ref{fig1}b). A short probe pulse excites the optical mode at $\omega_2= \omega_1+\Omega_{\rm m}$ and the response of the system is then recorded as a function of time (see Supplementary Information section 1). \textbf{b}, Time-domain measurements taken at the same set of cryostat temperatures as in Fig. \ref{fig2}b. Note that in these measurements, the probe frequency is centered at $\Omega_{\rm m}$, but has a large enough bandwidth to excite the optical mode even when it is detuned. \textbf{c}, Probe transmission as a function of time after probe pulse is turned on (top) and zoom in (bottom) showing oscillations at $\pi/g_{\rm m}$ and exponential decay with timescale $\tau \sim 2/\kappa$.
}
\label{fig3}
\end{figure}  

\newpage
\begin{figure}[]
\centerline{
\includegraphics[width=183mm]{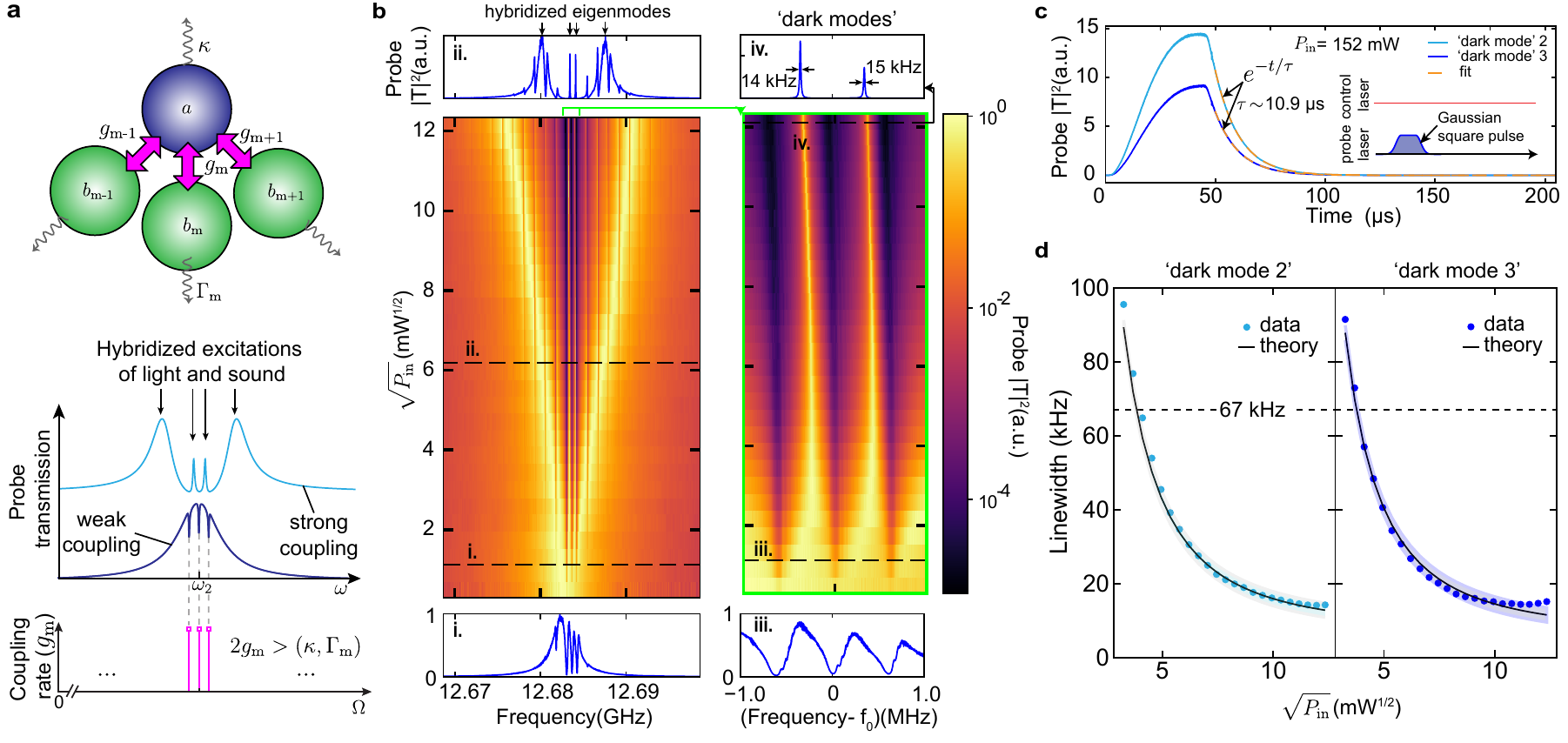}}
\end{figure}  
\captionof{figure}{ \textbf{Optomechanical strong coupling to three mechanical modes. } \textbf{a}, Diagram of linearized optomechanical coupling between an optical mode and three acoustic modes (top), corresponding expected spectra of probe laser transmission in the weak and strong coupling regime (middle), and the coupling rate under a strong laser drive (bottom). Strong coupling between an optical mode and three acoustic modes gives rise to four hybridized excitations of light and sound with two narrow resonances corresponding to optical `dark' modes and the two broad resonances corresponding to optical `bright' modes. \textbf{b}, Probe laser transmission spectra taken at various $P_{\rm in}$. The right panel shows a zoom-in of this spectra around frequency $f_o= 12.684 $ GHz. Lower and upper panels (inset \textbf{i-iv}) show spectra in the regimes of weak and strong coupling, respectively. In the strongly coupled case (insets \textbf{ii and iv}), two narrow resonances are observed, corresponding to the optomechanical dark modes. 
\textbf{c}, Time-domain measurement of dark modes. Inset shows the pulse sequence for the control and probe lasers. 
\textbf{d}, Measured linewidth of the two dark modes at various control laser powers and fits to theory with $\Gamma_i/2\pi = 5$ kHz as described in the main text. The shaded region corresponds to the theoretically-predicted linewidth of the two dark modes when we vary the fit parameter $\Gamma_i/2\pi$ from $3$ kHz to $7$ kHz.
}
\label{fig4}

\newpage

\clearpage


\pagebreak
\begin{center}
\textbf{\large Supplementary information for:\\ Multimode strong coupling in cavity optomechanics}
\end{center}

\setcounter{equation}{0}
\setcounter{figure}{0}
\setcounter{table}{0}
\setcounter{page}{1}
\setcounter{section}{0}
\makeatletter
\makeatletter \renewcommand{\thefigure}{S\@arabic\c@figure} \renewcommand{\thetable}{S\@arabic\c@table} \renewcommand{\theequation}{S\@arabic\c@equation}\makeatother 
\tableofcontents

\newpage
\section{Experiment} 
In this section, we describe the experimental apparatus used to perform frequency- and time-domain measurements of our optomechanical system. We first begin by discussing the experimental apparatus used to identify a pair of optical resonances necessary for phase-matched Brillouin interactions. 

\subsection{Identifying an optical mode pair:} \label{WaveSweep}
High-frequency acoustic modes mediate resonant inter-modal coupling between two distinct optical modes when energy conservation and phase-matching requirements are both satisfied (for details see Ref. \cite{kharel2018high}). Such requirements yield a characteristic `Brillouin frequency' $\Omega_\text{B} = 2\omega v_\text{a}/v_\text{o}$, where $\omega$ is the optical mode frequency and $v_\text{a}$ ($v_\text{o}$) is the velocity of sound (light) in the quartz ($z$-cut) crystal. For optical wavelengths near 1550 nm, we expect acoustic modes near $\Omega_\text{B} \simeq 2 \pi \times 12.6 $ GHz to mediate such phase-matched Brillouin interactions. However, since $v_\text{a}$ depends on the temperature of the quartz crystal, the Brillouin frequency changes as we cool our crystal from room temperature to cryogenic temperatures. Therefore, to precisely know Brillouin frequency as a function of wavelength, we determine $v_a$ from a prior measurement (see Ref. \cite{kharel2018high}) of the Brillouin frequency at cryogenic temperatures.   

\begin{figure}[]
\centerline{
\includegraphics[width=183mm]{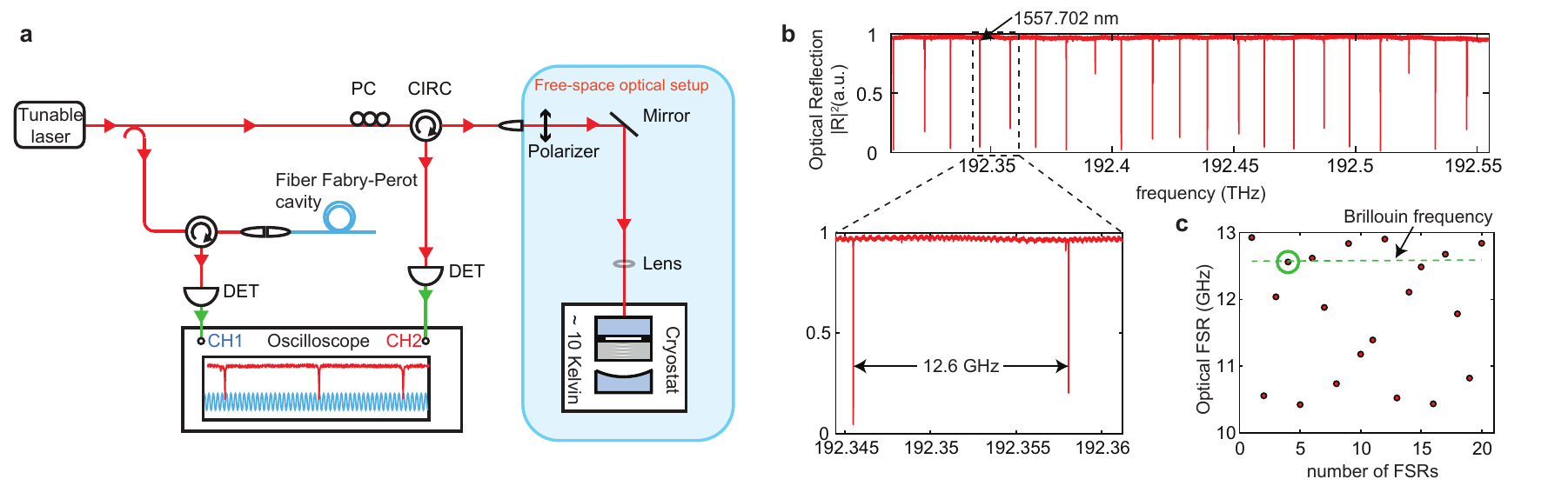}}
\caption{ \textbf{Finding an optical mode pair.} \textbf{a}, Schematic of the measurement apparatus used to determine resonant optical cavity modes within our system. Light from a continuously-tunable laser is coupled into the optical cavity using a combination of fiber and free-space optical components (PC: polarization controller; DET: detector; CIRC: circulator). Back-reflected light from the cavity is then recorded as a function of sweep time. Since the frequency sweep as a function of time is not perfectly linear, we use a separate fiber Fabry-P\'erot cavity with a known free spectral range to calibrate the frequency axis. \textbf{b}, Reflection spectrum measured at cryogenic temperature, by sweeping the tunable laser from 1556-1558 nm. Zoomed-in spectra showing two fundamental Gaussian optical modes with a free spectral range (FSR) of 12.6 GHz. \textbf{c}, We observed a large variation ($\sim$2.6 GHz) in the FSR with optical wavelength, permitting us to find multiple pairs of optical resonances having FSRs that match the Brillouin frequency.
}
\label{setup_FSR}
\end{figure}

To identify a pair of optical resonances that match the Brillouin frequency, we measure the reflection spectrum of the optical cavity using a setup as shown in Fig. \ref{setup_FSR}a. The output of a continuously-tunable laser (Agilent 8164B) is split into two arms. Light in one arm is coupled into the optical cavity through a fiber-optic collimator, a free-space polarizer, a mirror, and a lens. Light back-reflected from the optical cavity is collected using the same collimator, separated from the input light using a fiber circulator, and measured using a photodetector (Thorlabs PDA10CS2). Light in the other arm is coupled to a fiber Fabry-P\'erot (FFP) cavity formed by a segment of single mode fiber that is imperfectly coupled at one end. The back-reflected light from this FFP cavity is separated from the input light using a fiber circulator and detected using a photodetector. Since the wavelength sweep of our tunable laser is not perfectly linear as a function of time, we use the reflection spectrum of the FFP cavity as a frequency ruler for our swept measurements. This allows us to rapidly determine optical free-spectral ranges (FSR) ($\sim 13$ GHz) with an accuracy of $ \sim 1$ MHz for wavelength scans over several nanometers. 
From the reflection spectrum obtained by scanning the wavelength from 1556 to 1558 nm as seen in Fig. \ref{setup_FSR}b, we determine the optical FSRs as a function of laser frequency. Such measurements, an example of which is shown in Fig. \ref{setup_FSR}c, reveal a significant ($\sim 20 \%$) modulation in the FSRs. As discussed in Ref. \cite{kharel2018high}, such non-uniform optical mode spacing arises from optical reflections ($\sim 4\%$) in the quartz-vacuum interface. Due to the modulation in the FSR, we can readily find more than one pair of optical modes having a FSR close to the Brillouin frequency. Thermal tuning of the cryostat temperature using a proportional-integral-derivative (PID) controller (Cryocon 22C) then permits us to fine tune the optical FSR to exactly match the Brillouin frequency. 

\begin{figure}[]
\centerline{
\includegraphics[width=183mm]{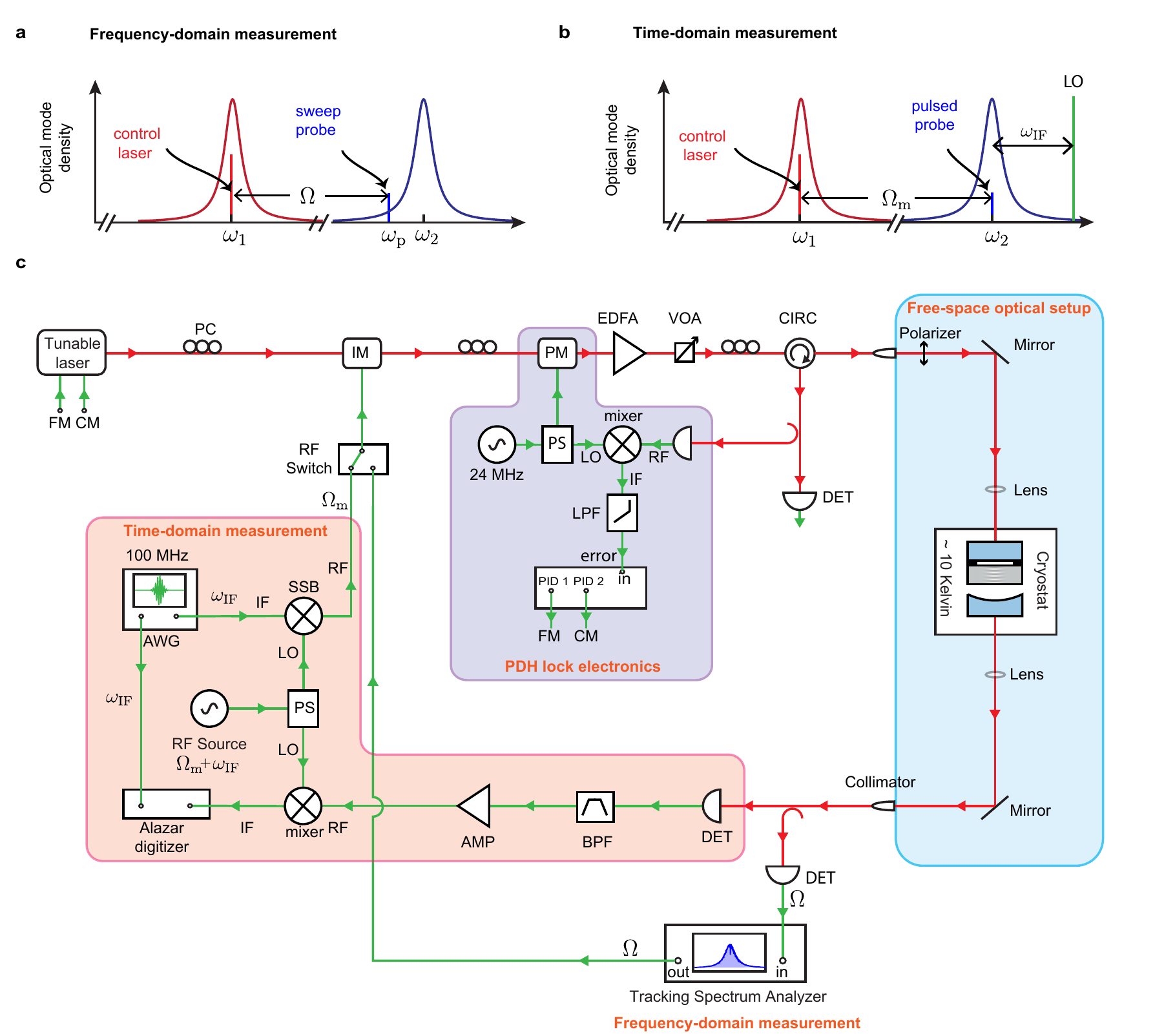}}
\end{figure}
\captionof{figure}{\textbf{Frequency-domain and time-domain measurement schematics.} \textbf{a}, For frequency-domain measurements, a strong control field is on-resonance with the low frequency optical mode ($\omega_1$) and a frequency tunable weak probe laser ($\omega_{\rm p} = \omega_1+\Omega$) is swept through the higher-frequency optical mode ($\omega_2$). \textbf{b}, For time-domain measurements, a strong control laser is continuously on resonance with the low frequency optical mode ($\omega_1$) but the weak probe laser ($\omega_{\rm p} = \omega_1+\Omega_{\rm m}$) is pulsed. This Gaussian-shaped probe pulse has a full-width-half-maximum (FWHM) of 10 ns. \textbf{c}, Schematics of the apparatus used to perform time-domain and frequency-domain measurements. RF: radio frequency; IF: intermediate frequency; LO: microwave local oscillator; AWG: arbitrary waveform generator, PS; RF power splitter; SSB: single-sideband mixer; AMP: RF amplifier; BPF: band-pass filter; LPF: low-pass filter, PC: polarization controller, IM: intensity modulator; PM: phase modulator; EDFA: erbium doped fiber amplifier; PID: proportional-integral-derivative controller; FM: frequency modulation; CM: current modulation;  VOA: variable optical attenuator.
}
\label{setup_fig}

\newpage

\subsection{Frequency-domain measurement:}
We use a well-known technique called optomechanically induced transparency (OMIT) to probe the coherent frequency response of our optomechanical system \cite{weis2010optomechanically}. This measurement is performed by using a control laser that is on-resonance with the lower-frequency optical mode at $\omega_1$ as seen in Fig. \ref{setup_fig}a. A probe laser synthesized from the same laser is swept through the higher-frequency optical mode at $\omega_2$. The transmitted probe light is measured as a function of probe laser detuning $\Omega=\omega_\text{p}-\omega_1$ using heterodyne detection.

In more detail, laser light from a tunable laser (Pure Photonics PPCL300) is locked on-resonance to the optical cavity mode at $\omega_1$ using the Pound-Drever-Hall (PDH) locking technique (See Fig. \ref{setup_fig}c). This light at $\omega_1$ is also intensity-modulated at a variable frequency $\Omega$ using a microwave signal generator (Agilent E8257D). This generates additional side-bands at $\omega_1 \pm \Omega.$  The tone at frequency $\omega_1$ serves as a strong control laser, whereas the weak tone at $\omega_1+\Omega$ serves as a probe laser. The side-band at $\omega_1-\Omega$ is irrelevant, as it is not resonant with any optical cavity mode due to the unequal FSRs. The intensity-modulated light is amplified using an erbium-doped fiber amplifier (EDFA) and coupled into the optical cavity using a fiber-optic polarization controller, a collimator, a free-space polarizer, a mirror, and a lens. Note that a fiber-optic variable optical attenuator placed after the EDFA is used to vary control laser power incident on the cavity. The light back-reflected from the cavity is separated from the incident light using a circulator. A part of this back-reflected light is used for the PDH lock.

Light transmitted through the optical cavity is collected using a free-space lens, a mirror, and a fiber-optic collimator and detected using a fast photoreceiver (Nortel Networks PP-10G). The spectrum analyzer monitors the heterodyne signal at $\Omega$ resulting from the beat-note of the transmitted probe laser and the transmitted control laser. The same spectrum analyzer also controls the frequency of the microwave signal generator, permitting us to track the heterodyne beat-note as a function of $\Omega$. 

\subsection{Time-domain measurement:}
To probe the time dynamics of our system, we use a control laser with a large, constant amplitude that is on-resonance with the optical mode at frequency $\omega_1$ and a weak probe pulse derived from the same laser that is on-resonance with the optical mode at $\omega_2$ as seen in Fig. \ref{setup_fig}b. Heterodyne detection of the probe signal transmitted through the cavity as a function of time is used to determine the time dynamics in our system.  

A weak excitation pulse is generated by intensity-modulating the laser output using a pulsed microwave drive at frequency $\Omega_\text{m}$ (See Fig. \ref{setup_fig}c); to generate such pulsed microwave drive, RF-output from a microwave signal generator (Berkeley Nucleonics 845-M) at $\Omega_\text{m}+\omega_\text{IF}$ is mixed with a pulsed output of an arbitrary waveform generator (Tektronix AWG5014C) at $\omega_\text{IF}$ using a single-sideband mixer. The control laser and the excitation pulse are both coupled into and out of the optical cavity through a combination of fiber-optic collimators and free-space optics described in the previous section. A part of the light transmitted through the cavity is detected using a fast photoreceiver, generating a RF beat-note at $\Omega_\text{m}$. This heterodyne signal, resulting from transmitted pump light beating with the transmitted probe light, is filtered, amplified, and demodulated using the RF-output (at $\Omega_\text{m}+\omega_\text{IF}$) of the microwave source. The demodulated RF-signal along with the reference RF-output of the AWG (both at $\omega_\text{IF}$) are recorded as a function of time using a digital oscilloscope (AlazarTech ATS9870). Comparing the demodulated signal with the AWG output allows us to measure both the phase and amplitude response of our optomechanical system.

\section{OMIT measurements at low power}

\begin{figure}[]
\centerline{
\includegraphics[width=183mm]{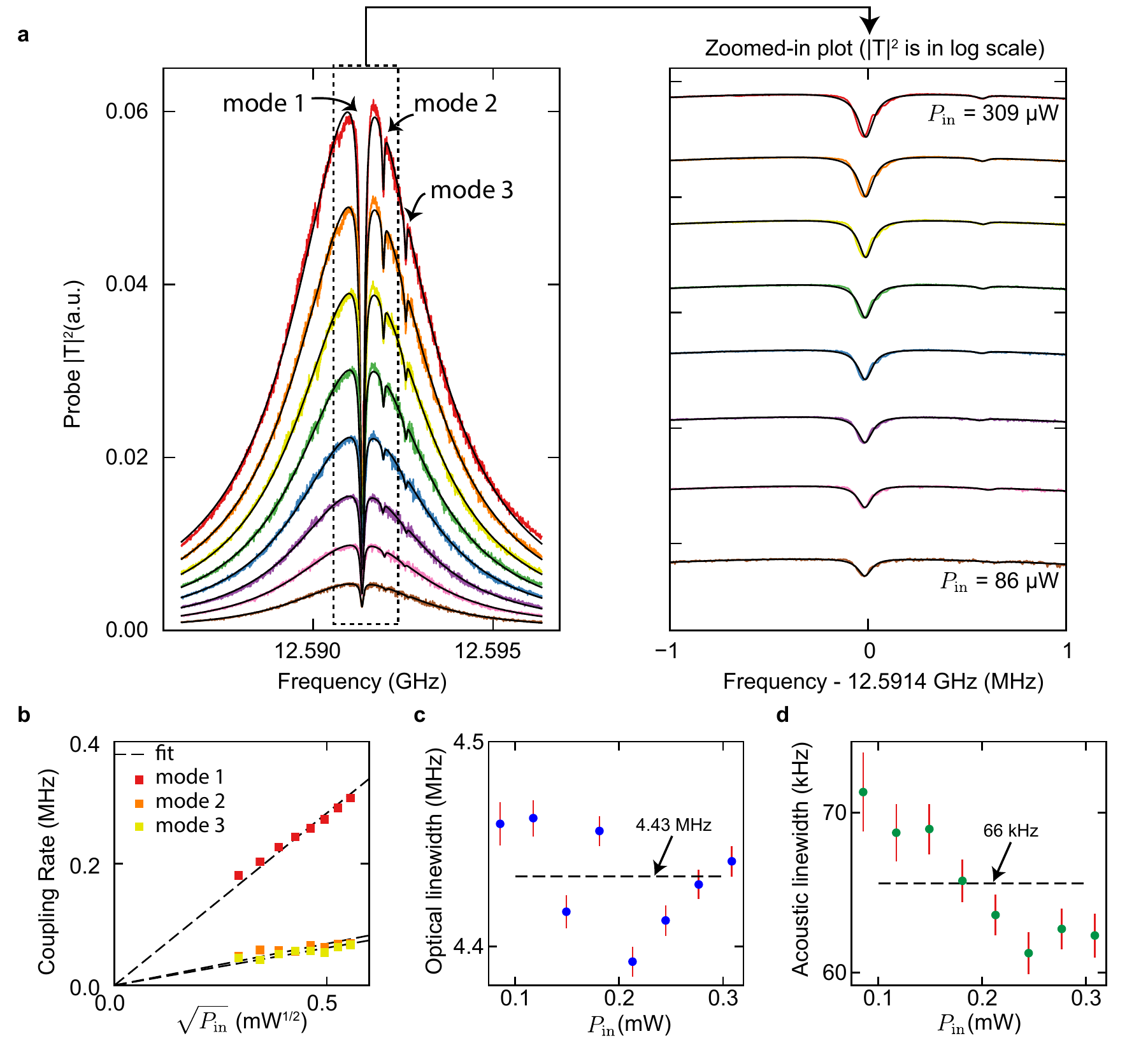}}
\caption{\textbf{OMIT measurement when coupled primarily to a single acoustic mode.} \textbf{a}, Probe transmission spectra obtained by varying input control laser power from 86 $\upmu$W to 309 $\upmu$W. Right panel shows a zoomed-in version of these OMIT spectra in log scale (each dataset is offset by 10 dB with respect to the previous for clarity). Theoretical fits to the spectra in \textbf{a} are used to obtain values of $g_{\rm m}$, $\kappa$, and $\Gamma_{\rm m}$ in \textbf{b, c}, and \textbf{d}, respectively. 
}
\label{OMIT_1mode}
\end{figure}
In this section, we describe the OMIT measurements performed at low input control laser powers to determine the optomechanical coupling rates as well as the optical and the mechanical damping rates. 

\subsection{Optomechanical coupling to one acoustic mode:}
The OMIT spectrum obtained through the heterodyne detection of the transmitted probe light ($a_\text{2, out}$) is given by \cite{kharel2018high} 

\begin{align} \label{OMITEq}
    P_{\rm d} (\Omega) \propto a_{2,{\rm out}}^\dagger a_{2,{\rm out}} =  \left|\frac{\kappa_{\rm ext}}{i(\Omega - \Delta)-\frac{\kappa}{2}+\sum_{m} \frac{g_{\rm m}^2}{-i(\Omega-\Omega_{\rm m}+\Gamma_{\rm m}/2)}} \right|^2,
\end{align}
where $\Omega= \omega_\text{p}-\omega_1$ is the pump-probe detuning, $\Delta= \omega_2 -\omega_1$ is the optical FSR, $g_{\rm m}$ is the coupling rate to acoustic mode at frequency $\Omega_\text{m}$ having a intrinsic dissipation rate of $\Gamma_\text{m}$, $\kappa$ ($\kappa_\text{ext}$) is the decay rate  (output coupling rate) of the optical mode at $\omega_2$. Since we use a symmetric Fabry-Perot optical cavity having two equal reflectivity ($99.9\%$) mirrors, the total optical cavity decay rate is dominated by the mirror's output coupling loss (i.e. $\kappa \simeq 2 \kappa_\text{ext}$). 

We determine $g_\text{m}$, $\Gamma_\text{m}$ and $\kappa$ by fitting experimentally obtained OMIT spectrum to Eq. (\ref{OMITEq}). Even though the optical mode weakly couples to many acoustic modes, OMIT spectra measured at low input control laser powers are described well by considering coupling to just 3 acoustic modes (See Fig. \ref{OMIT_1mode}a). From the values of $g_\text{m}$ obtained for the 3 acoustic modes (See Fig. \ref{OMIT_1mode}b) as a function of input control laser power $P_\text{in}$, we determine that the coupling strength to the acoustic mode directly on resonance (red) with the optical cavity mode is about 5 times larger than that for the other weakly coupled acoustic modes (yellow and orange). From fits to the OMIT spectrum (See Fig. \ref{OMIT_1mode}c,d), we also find that $\kappa= 2\pi \times (4.43 \pm 0.02)$ MHz and $\Gamma_\text{m} = 2 \pi \times (66 \pm 3)$ kHz. Note that we assumed equal intrinsic mechanical damping rate ($\Gamma_\text{m}$) for the three longitudinal acoustic modes having nearly equal frequencies. Since $g_\text{m} = \sqrt{\bar{n}_\text{c}} g_0^\text{m}$, we use the experimentally measured $g_\text{m}$ along with $\bar{n}_\text{c}$ determined from $P_\text{in}$ to obtain the single photon coupling rate $g_0^\text{m} = 2\pi \times (23 \pm 1) $ Hz. Note that $g_0^\text{m}$ obtained experimentally agrees well with with the theoretically predicted value of $\sim$ $2\pi \times 24$ Hz \cite{kharel2018high}.

\subsection{Optomechanical coupling to three acoustic modes:}
\begin{figure}[]
\centerline{
\includegraphics[width=183mm]{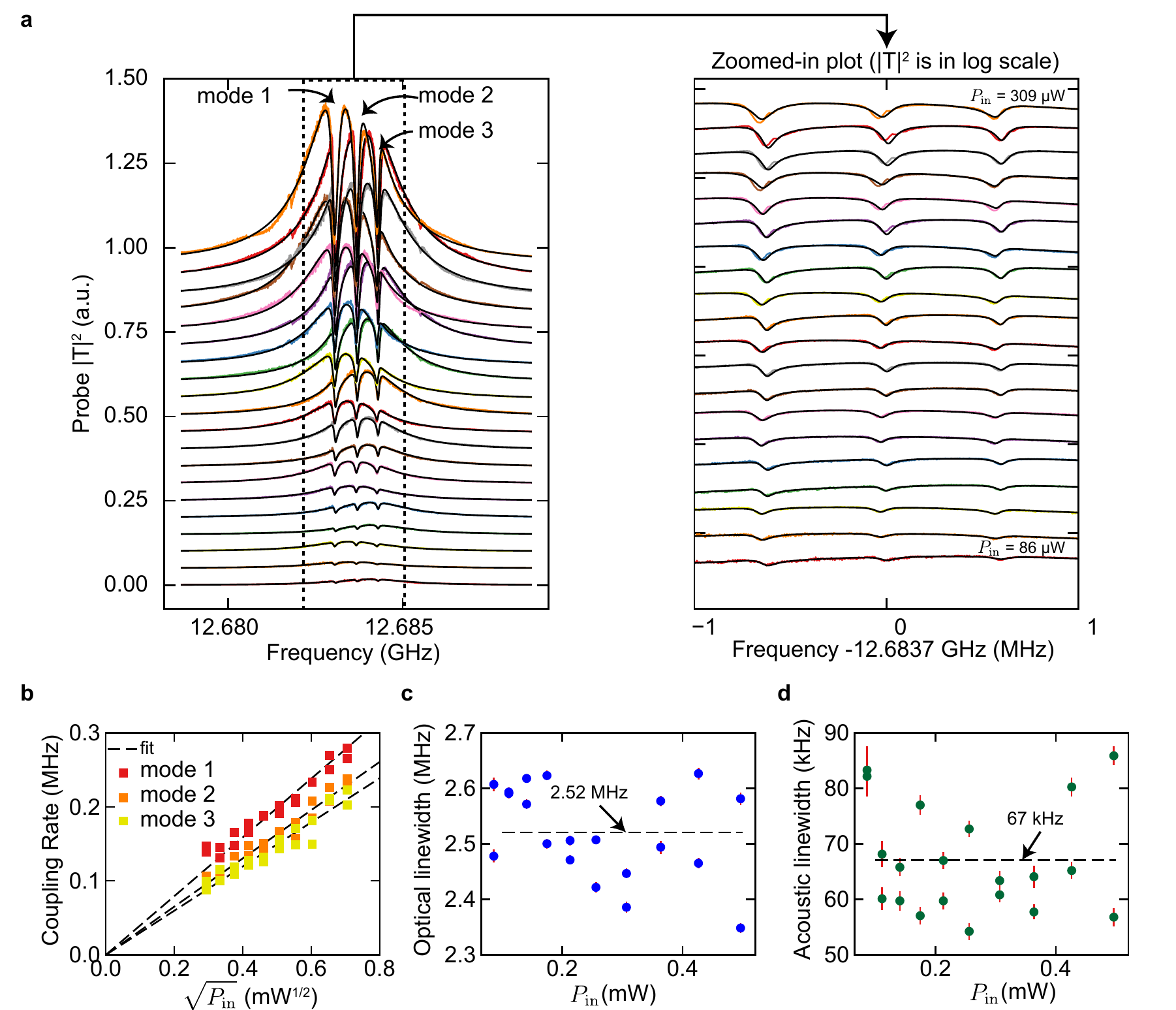}}
\caption{\textbf{OMIT measurement when coupled to three acoustic modes.} \textbf{a}, Probe transmission spectra obtained by varying input control laser power from 86 $\upmu$W to 496 $\upmu$W (each spectrum is offset from the previous by 0.05). The right panel shows zoomed-in OMIT spectra (each dataset is offset by 5 dB with respect to the previous for clarity). Theoretical fits to spectra in \textbf{a} are used to obtain values of $g_{\rm m}$, $\kappa$, and $\Gamma_{\rm m}$ in \textbf{b, c}, and \textbf{d}, respectively. 
}
\label{OMIT_3mode}
\end{figure}
It is possible to tailor the optomechanical coupling strength $g_\text{m}$ in our system. We engineer coupling to one or more acoustic modes near the Brillouin frequency by simply changing the optical wavelength. Optomechanical coupling to a given acoustic mode depends on the spatial overlap of the acoustic and optical modes. Even when the energy conservation and phase-matching requirements are satisfied for Brillouin interactions to occur, one can obtain vanishingly small optomechanical coupling in our system. For instance, this can occur when the spatial position of the crystal is such that the nodes of the acoustic mode profile lines up with the anti-nodes (nodes) of the optical forcing function (for more detail see supplementary information section II.D of Ref. \cite{kharel2018high}). Therefore, it is possible to vary the optomechanical coupling rate within our system either by changing the position of crystal inside the cavity or by changing the spatial profile of the optical forcing function. In absence of in situ control of the crystal position, we change the spatial profile of the optical forcing function by finding a different pair of optical modes (at a different wavelength) that still satisfy the Brillouin resonant condition described in section 1. Through this type of wavelength tuning, we can find a pair of optical mode such that inter-modal coupling is mediated simultaneously by three acoustic modes.

As in the previous section, we fit the experimentally obtained OMIT spectra as seen in Fig. \ref{OMIT_3mode}a  to determine coupling rates to these three acoustic modes near 12.684 GHz. Note that the three acoustic modes are separated by the acoustic FSR of $610 \pm 10$ kHz, which agrees well with the theoretically predicted FSR $v_\text{a}/2L_\text{ac} = 630$ kHz. A plot of the coupling rates as a function of $\sqrt{P_\text{in}}$ as seen in Fig. \ref{OMIT_3mode}b shows the expected linear dependence. This linear dependence can be used to extrapolate the following values for the coupling rates $g_\text{m}/2\pi$ at the highest input control laser power ($P_\text{in}$= 152 mW): $g_1= 2\pi \times (4.9 \pm 0.1)$ MHz, $g_2= 2\pi \times (4.0 \pm 0.1)$ MHz and $g_3= 2\pi \times (3.7 \pm 0.1)$ MHz. From the fits to the OMIT spectra, we find $\kappa=2\pi \times (2.52 \pm 0.08)$ MHz and $\Gamma_\text{m}= 2 \pi \times (67 \pm 10) $ kHz. Since $g_1,g_2,g_3> \kappa/2 \gg \Gamma_\text{m}/2$, this system is well-witin the multimode strong coupling regime.  

\section{Thermometry of acoustic mode}
One of the major challenges of using optomechanical systems for quantum information is ensuring that the mechanical resonator can be initialized in the quantum ground state. This challenge has been addressed by placing nano-scale GHz frequency resonators in a dilution refrigerator, or using the optomechanical interaction to cool the mechanical resonator. However, attempts to enhance the optomechanical coupling rate by increasing the control laser power eventually result in the system absorbing a significant amount of optical power, which leads to heating of the mechanical mode and increased intrinsic mechanical damping \cite{Chan2011}. These detrimental effects are the main limitations for state-of-the-art optomechanics experiments with nano-scale optomechanical resonators \cite{Riedinger2018}. We note that our device operates in a very different parameter space from those optomechanical systems and presents a different set of challenges and opportunities with respect to mitigating laser heating. 

Here we point out a few properties of our device that are significantly different from previously demonstrated optomechanical systems. Generally speaking, the dissipated optical power is proportional to the input power $P_{\rm in}$. We note that, while the circulating photon number $\bar{n}_c$ required to achieve strong coupling in our system is many orders of magnitude higher than those used in most experiments with nano-scale systems, $P_{\rm in}$ is not correspondingly larger by the same amount. This is due to the lower loss of our optical cavity and the fact our multimode optomechanical system allows our control laser to be on resonance with an optical mode rather than $\Omega_{\rm m}$ detuned. 

Another consideration is that the total dissipated power should be less than the cooling power of the cryostat and therefore does not heat up the entire system. Beyond this basic requirement, however, the fraction of power absorbed by the mechanical resonator itself needs to be low enough so that the acoustic mode of interest remains in the ground state and maintains its coherence properties. In most nano-scale systems, the fraction of dissipated power that is absorbed by the optomechanical resonator chip itself is significant. In our system, as we will discuss below, the optical dissipation is likely dominated by light elastically scattering out of the optical mode and being absorbed elsewhere, while the acoustic resonator crystal itself experiences very little absorption. Furthermore, the mass of our acoustic resonator is 1-100 million larger than those of typical micro/nano-scale optomechanical systems, resulting in an even smaller change in temperature. Therefore, we believe that the main challenge in our system will be reducing the total dissipated power by, for example, increasing the cavity finesse (therefore reducing $P_{\rm in}$) and reducing scattering losses. 

As a first step toward quantifying the effects discussed above, we now present thermometry measurements of our optomechanical system and a study of the effects of laser heating.

\begin{figure}[]
\centerline{
\includegraphics[width=183mm]{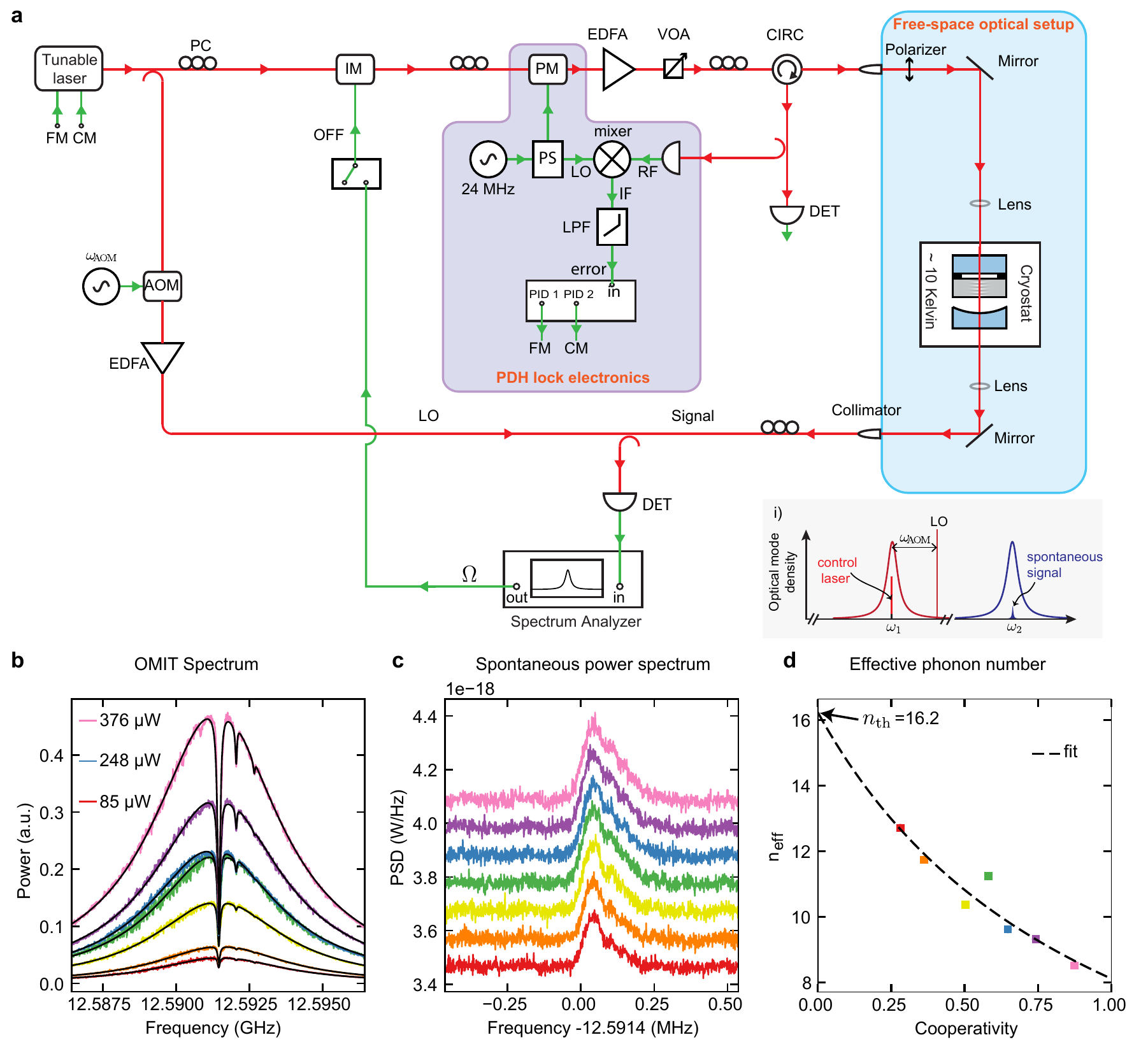}}
\caption{ \textbf{Measuring $n_{\rm th}$.} \textbf{a}, Schematic of the apparatus used to measure the spontaneously-scattered light due to thermal fluctuations of the mechanical mode. \textbf{b}, OMIT spectra and \textbf{c}, spontaneous spectra at various control laser powers. \textbf{d}, Extracted effective phonon number $n_{\rm eff}$ as a function of $C$. From a theoretical fit to this data, we obtain $n_{\rm th} \approx 16.$ 
}
\label{spont_1mode}
\end{figure}

\subsection{Determining $\bar{n}_\text{th}$:}
In this subsection, we determine the thermal occupation number ($\bar{n}_\text{th}$) by performing calibrated measurements of spontaneously-scattered light due to thermal fluctuations of the mechanical mode. For such measurements, a weak control laser is placed on resonance with the lower-frequency mode at $\omega_1$ (See Fig.  \ref{spont_1mode}a). Unlike stimulated measurements (or OMIT measurements) discussed in section S1.2, no input probe light is supplied (i.e. the microwave source driving the intensity modulator, which generates the probe side-band, is turned off). The thermally-populated acoustic mode at frequency $\Omega_\text{m}$ then spontaneously scatters light to frequency $\omega_1+\Omega_\text{m}$ through the anti-Stokes process (See inset i. of Fig.  \ref{spont_1mode}a). This spontaneously scattered light ($a_2$) exits the optical cavity through both the reflection and the transmission port. The transmitted light is collected and measured using heterodyne detection. An acousto-optic modulator frequency shifts the control laser to generate an optical local oscillator (LO) at frequency $\omega_1+ \omega_\text{AOM}$ for this detection scheme. This LO is amplified using an erbium-doped fiber amplifier (EDFA) to enhance signal-to-noise in our spontaneous measurements. 

The power spectrum of the spontaneously scattered light ($a_2$) is given by \cite{kharel2018high}
\begin{align} \label{PSpectrum}
    S_{a_2a_2}(\Omega)\simeq \frac{g_\text{m}^2 \Gamma_\text{m} \bar{n}_\text{th}}{(\kappa/2)^2 \left((\Omega-\Omega_\text{m})^2+(\Gamma_{\rm m}^\text{eff}/2)^2\right)}
\end{align}
where $g_{\rm m}$ is the optomechanical coupling rate, $\Gamma_\text{m}$ is intrinsic mechanical decay rate, $\kappa$ is the linewidth of the optical mode at $\omega_2$ and $\Gamma_{\rm m}^\text{eff}$ is the effective damping rate of the acoustic due to optomechanical coupling. $\Gamma_\text{m}^\text{eff} = \Gamma_\text{m} (1+C)$, where the optomechanical cooperativity $C= 4 g_{\rm m}^2/(\kappa \Gamma_\text{m}).$ Note that we assumed $\kappa \gg \Gamma_\text{m}$ while deriving Eq. (\ref{PSpectrum}). The total intra-cavity spontaneous scattering rate due to thermal fluctuations is then given by
\begin{align}
\Gamma_\text{opt}= \frac{1}{2\pi} \int_{-\infty}^{\infty} \text{d}\Omega \ S_{a_2a_2} = \frac{4g_\text{m}^2 \Gamma_\text{m} \bar{n}_\text{th}}{\kappa^2 \Gamma_\text{m}^\text{eff}}.
\end{align}
Since $a_\text{2,out} =\sqrt{\kappa^\text{ext}} a_2$, the total optical power ($P_{\rm t}$) in the spontaneously scattered signal transmitted through the optical cavity is
\begin{align}
    P_{\rm t} = \hbar \omega \kappa^\text{ext} \Gamma_\text{opt}
\end{align}
The total spontaneously scattered power at the detector $P_{\rm d}$ = $\eta P_{\rm t}$, where $\eta$ is the system detection efficiency resulting from optical losses and detector response. The total spontaneously scattered power can be written in terms of the effective acoustic number $\bar{n}_\text{eff} = \Gamma_m \bar{n}_\text{th}/\Gamma_\text{m}^\text{eff}$ as 
\begin{align} \label{eqneff}
    P_\text{d} = \eta P_\text{t} = \frac{4\eta \hbar \omega \kappa^\text{ext} g_\text{m}^2}{\kappa^2} \frac{\Gamma_m n_\text{th}}{\Gamma_\text{m}^\text{eff}} = \frac{4 \eta \hbar \omega \kappa^\text{ext} g_\text{m}^2}{\kappa^2} \bar{n}_\text{eff}.
\end{align}
Since $\bar{n}_\text{eff} =\Gamma_\text{m} \bar{n}_\text{th}/\Gamma_\text{m}^\text{eff} = \bar{n}_\text{th}/(1+C)$, $\bar{n}_\text{eff} $ decreases as we increase $C$ by increasing the input control. Therefore, increasing the control laser power effectively cools the acoustic mode. 

We determine $P_\text{d}$ through spontaneous measurements and $g_{\rm m}$, $\kappa$, and $\Gamma_{\rm m}$ through OMIT measurements (See Fig. \ref{OMIT_1mode} b,c). After measuring $\eta$, we determine $n_{\rm eff}$ as a function $C$. Note that as a first-order approximation we assumed $\kappa_{\rm ext}=\kappa/2$ (See Fig. \ref{OMIT_1mode} d). A fit to $n_{\rm eff}$ as a function to $C$ then allows us to determine $n_{\rm th} \approx 16$. This occupation number (for a acoustic mode at 12.6 GHz) corresponds to a bath temperature of $\sim$ 10 K, which is consistent with the operation temperature of $\sim$ 12 K. After performing this measurement, however, we determined that accurate measurement of thermal occupation number requires precise measurement of the external coupling rate for the optical mode at $\omega_2$. The addition of the crystal inside our Fabry-Perot optical cavity leads to asymmetry in the external coupling rates for the input and output mirrors and also introduces additional intrinsic loss channel for the optical cavity modes, which were not taken into account in the above measurement. Therefore, in the following sections we explore this asymmetry in the external coupling rates and perform further thermometry measurements to obtain a more accurate determination of the phonon occupation number $n_{\rm th}$. 



\begin{table}
  \centering
  \resizebox{\columnwidth}{!}{
  \begin{tabular}{|c|c|c|c|c|c|c|}
    \hline 
    Optical mode & $\kappa/2\pi$ (MHz) &$\kappa_1^{\rm ext}/2\pi$ (MHz) & $\kappa_2^{\rm ext}/2\pi$ (MHz) & $\kappa^{\rm in}/2\pi $(MHz) & Norm. Reflection & Norm. Transmission \\
    \hline
    i) & 2.47 & 0.81 & 1.21 & 0.45 & 0.12 & 0.64 \\
    \hline
    ii) & 2.42 & 1.04 & 1.10 & 0.27 & 0.02 & 0.79 \\
    \hline
    ii) & 2.35 & 1.00 & 1.10 & 0.25 & 0.02 & 0.80\\
    \hline
  \end{tabular}
  }
  \caption{{\bf Internal and external loss rates.} Measured internal and external losses, as well as fraction of light reflected and transmitted on-resonance for three optical modes shown in the bottom panel of Fig. \ref{Internal_loss_and_heating}c.} \label{table:1}
\end{table}

\subsection{Internal losses and asymmetric coupling rate:} \label{intloss}

\begin{figure}[]
\centerline{
\includegraphics[width=165mm]{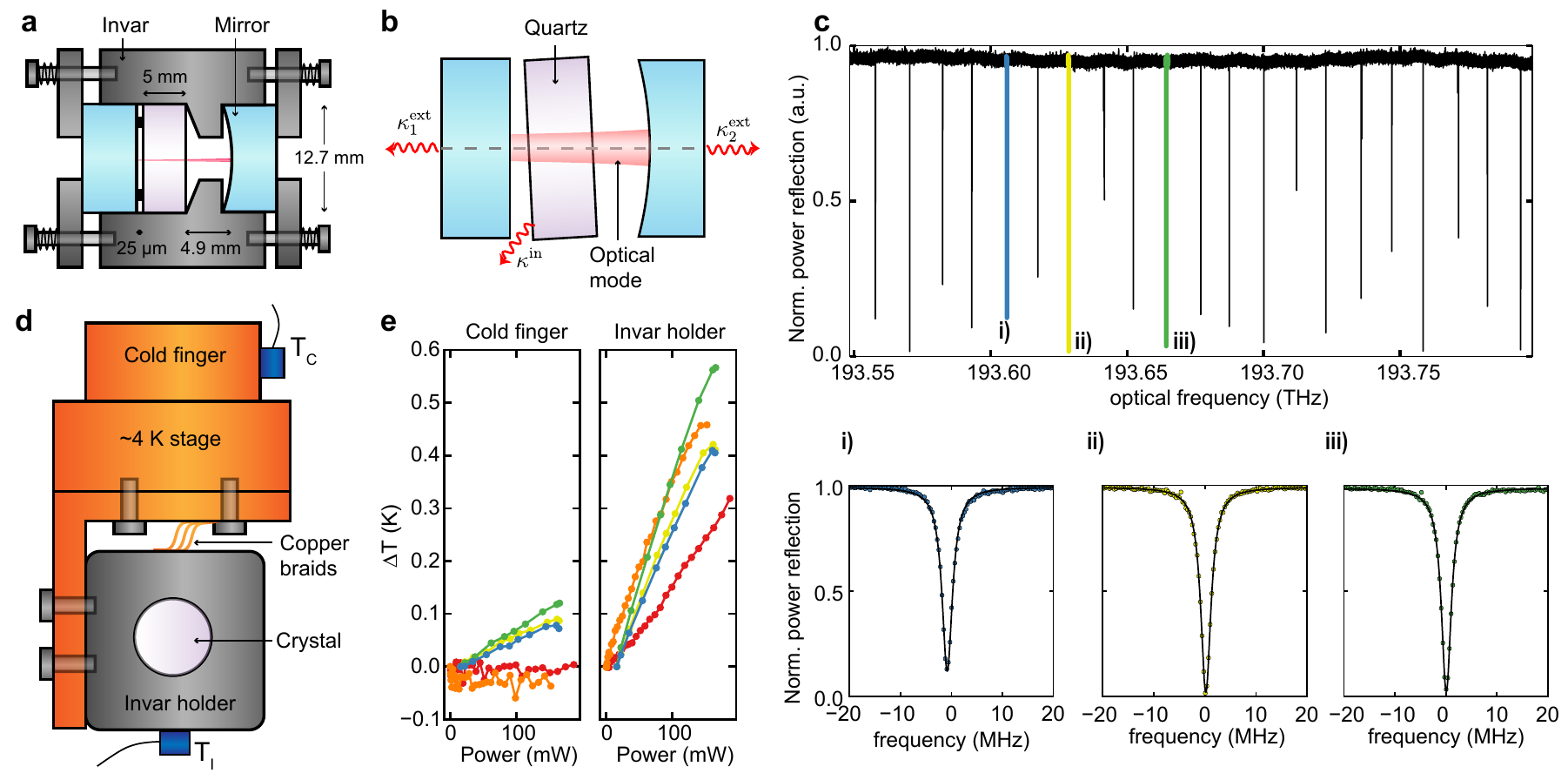}}
\caption{\textbf{Measurement of internal losses and asymmetric external coupling rate.} \textbf{a}, Schematic of the plano-concave optical cavity and quartz crystal housed in a Invar holder. \textbf{b}, The placement of a quartz crystal inside the optical cavity introduces an asymmetry in the external coupling rate as well as additional internal losses due to misalignment or absorption. \textbf{c}, Top panel shows the measured reflection spectrum obtained by scanning a tunable laser source from 1548-1550 nm. Bottom three panels shows zoomed-in spectra for three different optical modes. \textbf{d}, Schematic depicting points at which we measure the temperature of the cold finger ($T_c$) and the temperature of the Invar holder ($T_I$). \textbf{e}, Measurements of the temperature changes $\Delta T_c$ and $\Delta T_I$ as we vary the incident optical power.}
\label{Internal_loss_and_heating}
\end{figure}

In this section, using a combination of reflection and transmission measurements, we determine the internal loss rate as well as asymmetry in the external coupling rates for optical modes within our nearly hemispherical Fabry-P\'erot cavity. 

Our plano-concave optical cavity consists of two equal reflectivity (99.9\%) mirrors with a quartz crystal that is placed in between the mirrors (see Fig. \ref{Internal_loss_and_heating}a). The introduction of the crystal can give rise to additional loss mechanisms for optical modes. These include bulk/surface absorption of light in quartz and scattering losses due to misalignment of the crystal axis relative to the optical cavity axis (see Fig. \ref{Internal_loss_and_heating}b). $\kappa^{\rm in}$ characterizes all such internal loss mechanisms. Moreover, small optical reflections ($\sim $4\%) in the quartz-vacuum interfaces and the positioning of the crystal within the cavity can give rise to a difference in the external coupling rates for the input ($\kappa_1^{\rm ext}$) and output {$\kappa_2^{\rm ext}$} mirrors even when the mirror reflectivities are equal (see Ref. \cite{kharel2018high}).

To quantify $\kappa^{\rm in}$, $\kappa_1^{\rm ext}$, and $\kappa_2^{\rm ext}$ at cryogenic temperatures, we first measured the optical reflection spectrum. As explained in section \ref{WaveSweep}, this measurement is performed by sweeping the frequency of light incident on the cavity and recording the back-reflected light’s power (see Fig. \ref{Internal_loss_and_heating}c). Next, we measured the fraction of light transmitted on resonance for several optical modes labelled i), ii), iii) in Fig. \ref{Internal_loss_and_heating}c).

The input-output formalism \cite{walls2007quantum} of an asymmetric Fabry-P\'erot cavity gives the following expressions for the reflection $R(\omega)$ and transmission spectrum $T(\omega)$ of an optical mode at $\omega_0$

\begin{align}\label{Romega}
R(\omega) &= \left| \frac{i(\omega-\omega_0)-\kappa/2 +\kappa_1^{\rm ext}}{\kappa/2-i(\omega-\omega_0)}\right|^2, \\ \label{Tomega}
T(\omega) &= \left|\frac{\sqrt{\kappa_1^{\rm ext}\kappa_2^{\rm ext}}}{\kappa/2-i(\omega-\omega_0)}  \right|^2.
\end{align}

By fitting the reflection spectra seen in Fig. \ref{Internal_loss_and_heating}c.i-iii to Eqn. (\ref{Romega}), we determine $\kappa$ and $\kappa_1^{\rm ext}$. We then use the fraction of light transmitted on resonance ($T(\omega_0)$) to determine $\kappa_2^{\rm ext}$. Finally, we use $\kappa = \kappa^{\rm in}+\kappa_1^{\rm ext}+\kappa_2^{\rm ext}$ to obtain $\kappa^{\rm in}$. Table \ref{table:1} shows measurement of $\kappa^{\rm in}$, $\kappa_1^{\rm ext}$, and $\kappa_2^{\rm ext}$ for three different modes. On average, we measured $\kappa^{\rm in}/\kappa \sim 0.13 \ll 1$. Also notice that the external coupling rates can differ by as much as 34\% relative to $\kappa/2$ (see optical mode i in Table \ref{table:1}). Interestingly, even during a single cryogenic experimental run, we observed variation in internal loss rates for different optical modes. This, for example, could occur when part of the internal loss is due to crystal misalignment or surface absorption as optical modes having the nodes (anti-nodes) at the quartz-vacuum interfaces are less (more) sensitive to such scattering losses at the interfaces. 

Prior measurements of optical absorption in quartz have shown bulk absorption coefficient as low as 4.3 dB/km (at 1.06 $\upmu$m wavelength), which would correspond to an internal loss rate of $\kappa^{\rm in}/2\pi \sim 20$ kHz in our optomechanical system. It is interesting to note that the authors in Ref. \cite{pinnow1973development} performed calorimetric measurement of optical absorption in quartz using a geometry that is similar to our system (i.e., by placing a quartz crystal inside a laser cavity). Therefore, by modifying the Invar holder design to allow in situ alignment and improving the crystalline resonator geometry (for example using a plano-convex crystal) to minimize scattering losses, we should be able to increase the finesse of our optical cavity to at least $>3\times 10^4$ from our present values of $\sim 5 \times 10^3$ by using higher reflectivity mirrors. 

The internal losses within our optical cavity lead to a small but non-negligible increase in temperature of the Invar holder $T_{\rm I}$ and the cold-finger $T_{\rm c}$ (see Fig. \ref{Internal_loss_and_heating}d). To study such ‘laser heating’ effects, we monitored $T_{\rm I}$ and $T_{\rm c}$ by varying the power of the incident control laser that is on resonance with different cavity modes over multiple cooldown runs. Through these measurements, we observed an average temperature change of 3 mK (0.4 mK) of the Invar-holder (cold-finger) per 1 mW of incident input control laser power (see Fig. \ref{Internal_loss_and_heating}e). Note that a negligible temperature change was observed when the control laser was not on-resonance with the optical mode, suggesting that scattering and absorption of light inside the optical cavity are dominant sources of laser heating.


\subsection{Laser heating:}

\begin{figure}[]
\centerline{
\includegraphics[width=165mm]{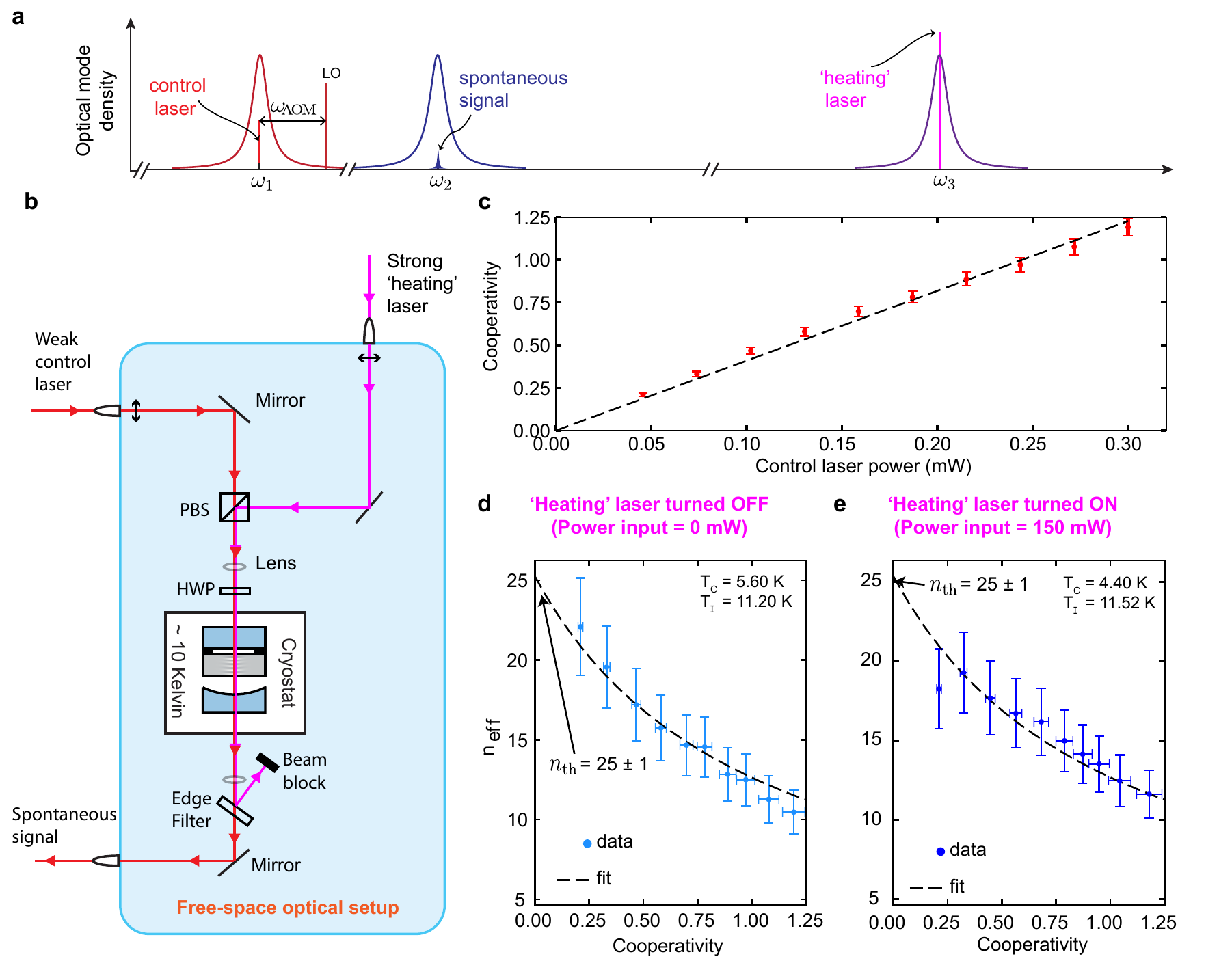}}
\caption{\textbf{Optomechanical thermometry in presence of a heating laser.} \textbf{a}, Schematic spectra of optical modes showing a weak control laser that is resonant with the optical mode at $\omega_1$ and a strong laser at $\omega_3$ to heat up the cavity. \textbf{b}, Schematic of the free-space optical setup used to couple the weak control laser and the strong heating laser into the optical cavity. PBS: polarizing beam splitter; HWP: half-wave plate. \textbf{c}, Measurement of $C$ obtained by fitting OMIT dips as a function of input control laser power $P_{\rm in}$. \textbf{d}, With the heating laser turned off, we measured $\bar{n}_{\rm th} = 25.2\pm1.2$, corresponding to a bath temperature of $15.6\pm0.7$ K for 12.6 GHz phonons. \textbf{e}, With the heating laser turned on (150 mW input), we still measured $\bar{n}_{\rm th} = 25.4\pm1.1$, corresponding to a bath temperature of $15.7\pm 0.7$ K. }
\label{thermometry}
\end{figure}

In this section, we perform further experiments to explore how laser heating could affect the performance of our optomechanical system. In particular, we explore how such heating changes the thermal decoherence rate of bulk acoustic modes within the quartz crystal and whether it is feasible to operate our optomechanical system at temperatures below 1 K within a standard dilution refrigerator.

We performed optomechanical thermometry measurements with and without a separate ‘heating laser’ to explore changes in the thermal decoherence rate ($\gamma_{\rm m} = \bar{n}_{\rm th} \Gamma_{\rm m}$) for the mechanical mode  (see Fig. S7a). As explained before, we determine $\bar{n}_{\rm th}$ by performing calibrated measurements of spontaneously-scattered light resulting from thermal fluctuations of the mechanical mode. These measurements are performed at low input powers $P_{\rm in}<$0.4 mW because even though we couple predominantly to a single acoustic mode, the effect of a multitude of weakly-coupled acoustic modes on the OMIT spectra cannot be ignored at high powers. In addition, the dependence of $n_{\rm eff}$ on $P_{\rm in}$ becomes more complicated as we approach strong coupling, and cooling becomes less efficient \cite{MarquardtPRL2007}. However, we routinely use $P_{\rm in}$ as large at 170 mW to push our system well in the strong coupling regime. Therefore, measurements of effective acoustic number at $C<1$ might not be adequate to quantify the effects of laser heating. Therefore, to understand how such large incident optical power could affect the bath temperature of our acoustic mode, we used a separate heating laser in addition to the weak control laser (at 1552.453 nm wavelength) used to perform thermometry. The heating laser was on resonance with a nearby optical cavity resonance (at 1528.724 nm wavelength). We use a polarizing beam splitter to combine the control laser and the heating laser before coupling it to the optical cavity. A free-space high-pass filter having a transmission edge at $\sim$1538 nm is used to reject the strong heating laser before collecting the weak spontaneously-scattered signal for sensitive heterodyne detection so as to not saturate the photodetector. 

We first measured the effective thermal occupation number of the acoustic mode as a function of the cooperativity with the heating laser turned off. As before, for each $P_{\rm in}$, we recorded both the OMIT spectra and the spontaneously scattered signal. We also measured $\eta =0.37 \pm 0.05$ and $\kappa_2^{\rm ext}= 2\pi \times (0.92\pm 0.03)$ MHz to obtain a calibrated measurement of $n_{\rm eff}$ as a function of $C$ (See Fig. c-d). Accounting for the asymmetry in the external coupling rate, we obtained a more accurate measurement of $\bar{n}_{\rm th}  = 25.2 \pm 1.2 $. This $\bar{n}_{\rm th}$ corresponds to a bath temperature of $15.6\pm 0.7 $K for the 12.6 GHz acoustic mode. Note that this temperature is slightly higher than the independently measured Invar temperature of $11.2$ K suggesting a non-negligible thermal resistance between the quartz and the Invar holder. Finally, we obtained $\Gamma_{\rm m} = 2 \pi \times (61\pm 3)$ kHz from the OMIT measurements, which gives $\gamma_{\rm m} = 2\pi \times (1.6\pm 0.1) $ MHz at $15.6\pm 0.7 $K.

Next, we turn on the heating laser (input power $\sim150$ mW) and perform optomechanical thermometry of the same acoustic mode. As discussed in section \ref{intloss}, adding a strong heating laser into the optical cavity increased the temperature of the Invar holder by 0.78 K. Similar to what was shown in Figure 2b of the main text, a change in temperature leads to a change in the cavity FSR of approximately 6 MHz, which is consistent with the linear thermal expansion of the Invar holder at cryogenic temperatures. However, to perform thermometry experiments we need to compensate this detuning between the optical FSR to the Brillouin frequency. To do this in absence of independent in situ control of the optical cavity length, we slightly changed the setpoint of the PID temperature control so that $T_{\rm I} = 11.52$ K. We note that this adjustment in temperature ($\Delta T_{\rm I} = 0.32$ K) is within the uncertainty of our extracted bath temperatures, and this should not significantly affect the conclusions of the measurement. From the measurement of $n_{\rm eff}$ as a function of $C$, we obtained a thermal occupation number of $25.4 \pm 1.1 $ even in presence of a strong heating laser. This thermal occupation corresponds to a bath temperature of $15.7 \pm 0.7$ for the 12.6 GHz acoustic mode. Moreover, we observed that $\Gamma_{\rm m} = 2 \pi \times (64\pm 3)$ kHz, corresponding to $\gamma_{\rm m} = 2\pi \times (1.6\pm 0.1) $ MHz. Within the uncertainty of our measurement, we observe that adding a strong laser light does not alter the acoustic decoherence rate.

These measurements indicate that at 10 K, our system is in the quantum coherent regime and the acoustic modes are robust to thermal decoherence from laser heating.

\subsection{Prospects for ground state operation:}
We now discuss the feasibility of reaching the strong coupling regime within a dilution refrigerator. For continuous wave (CW) operation, the cooling power of the refrigerator has to be at least larger than the optical heating rate.  Increasing the optical finesse to $3\times 10^4$ from the present value of $5\times 10^3$ would allow us to reach the regime of strong coupling at $<100 \ \upmu$W input optical powers.  
Since the optical heating rate cannot be larger than the incident optical power, a dilution refrigerator operating at $\sim$ 200 mK could then handle such optical heating rates of the order of 100 $\mu$W even in the worst-case scenario that all of the optical power is dissipated in the system. However, since pulsed operation is more desirable for a variety of quantum operations such as such as a state swap during quantum transduction or quantum information storage, we expect the heat load for such experiments to be at least an order of magnitude smaller than that for the the CW operations. Note that the estimates that we made are conservative and, in principle, it is possible to reduce such intrinsic loss mechanisms through improvements in the optical cavity, crystalline resonator design and choosing various pristine crystalline solids such as CaF$_2$ and diamond. 

In summary, we have shown that our optomechanical device as a whole has the potential to be relatively robust to laser heating. With modest improvements to the geometry and experimental parameters, it can be a promising system for ground state operation and quantum state manipulation of bulk acoustic phonons.

\section{Frequency and time-domain correspondence}

\begin{figure}[ht]
\centerline{
\includegraphics[width=183mm]{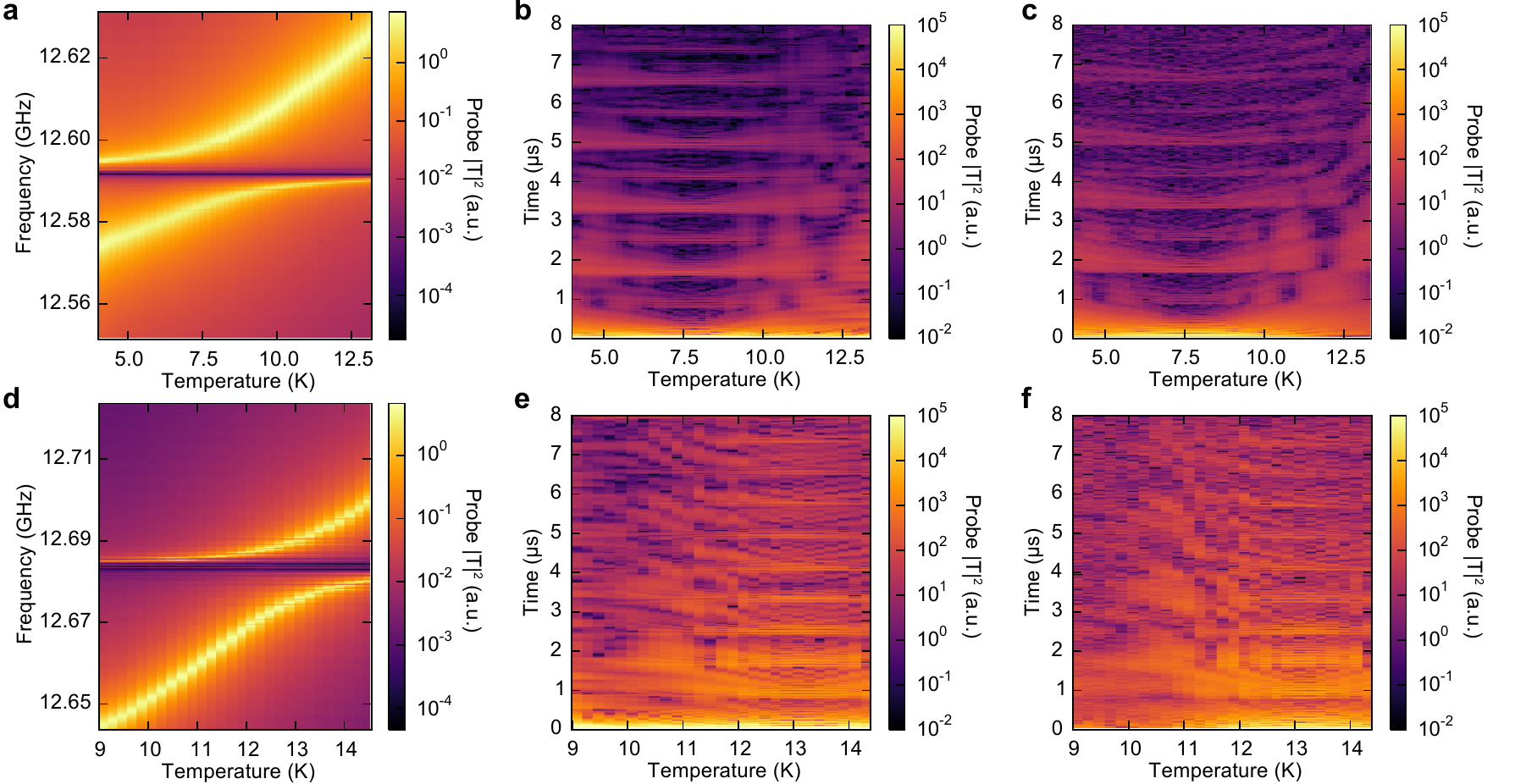}}
\caption{ {\bf Comparison of time-and frequency-domain measurements.} \textbf{a)}, Temperature dependence of the probe transmission spectrum \textbf{b)}, its Fourier transform \textbf{c)}, and time-domain data for the case of one strongly coupled mode. \textbf{d, e}, and \textbf{f} are the same for the case of three strongly coupled modes. The Fourier transforms have been scaled to roughly match the amplitude range of the time-domain data.
}
\label{FFT_t}
\end{figure}

In Figure \ref{FFT_t}, we present the temperature dependent frequency and time-domain data for both one (\textbf{a-c}) and three strongly coupled acoustic modes (\textbf{d-f}). In addition, we show the Fourier transform of the frequency-domain data, which have the same qualitative features as the time-domain data. This indicates that the intricate time dynamics we observe are due to the multimode acoustic spectrum observed in the frequency-domain. For example, we see that in the case of three strongly coupled modes, the revivals are much more prominent than in the one-mode case. 

\section{Dissipation rates in multimode strong coupling}

In this section, we provide a theoretical description of our optomechanical system in the multimode strong coupling regime. We first show how to include the effects of interference when both acoustic modes decay to a common reservoir  \cite{CarmichaelPRL1993, metelmann2015nonreciprocal}. Using the simplest case of two acoustic modes and one optical mode, we review how strong coupling gives rise to a optomechanical ``dark mode", which have been described in previous works \cite{massel2012multimode}. We then derive a analytic expression for the dark mode's dissipation rate. This simple result illustrates the physical intuition behind the novel phenomenon of line-narrowing observed in our experiment.

In the rotating frame of the pump, the undriven and linearized Hamiltonian for our system under the rotating wave approximation is given by
\begin{align}
    H &= \Delta a^{\dagger}a + \sum_{m}\Omega_m b_m^{\dagger}b_m +\sum_{m} g_m(a^{\dagger}b_m + b_m^{\dagger}a)
\end{align}

\begin{figure}[]
\centerline{
\includegraphics[width=80mm]{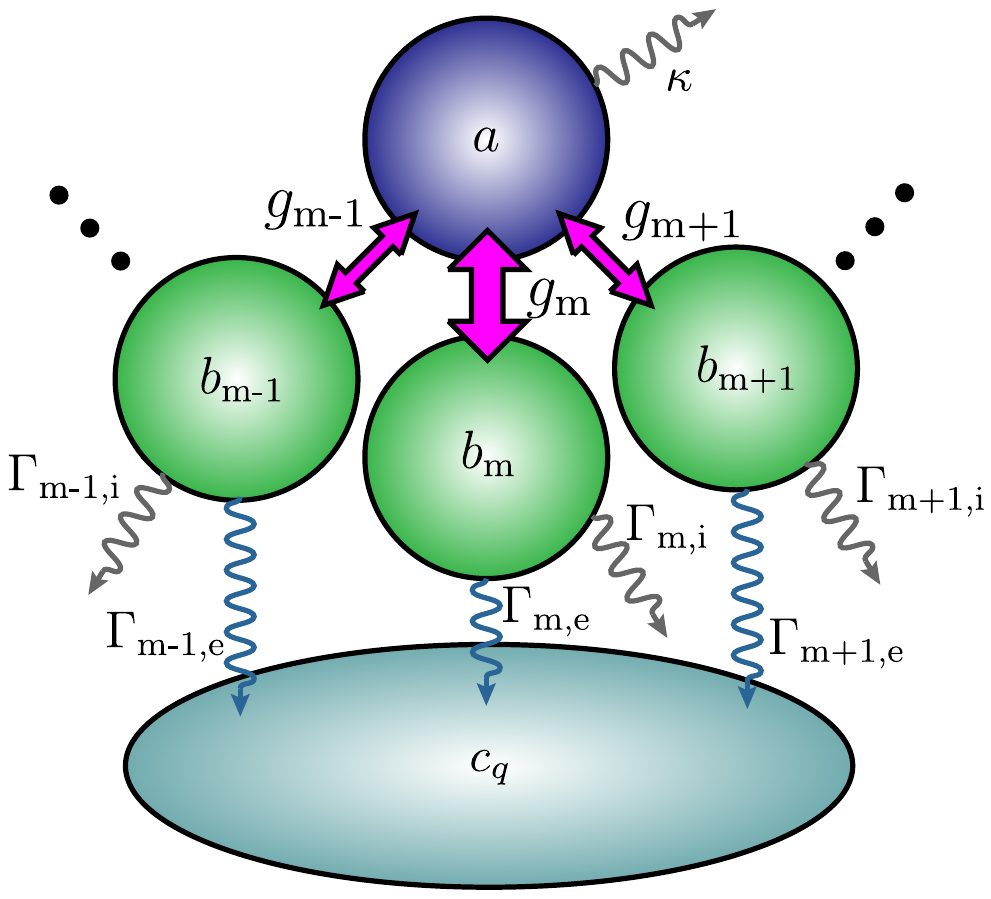}}
\caption{\textbf{Multimode cavity-optomechanical.} Dissipative coupling arises when multiple acoustic modes decay into a common bath.
}
\label{SDispCartoon}
\end{figure}  

We include the effects of dissipation in our system using the model illustrated in Figure \ref{SDispCartoon}. As discussed in the main text and in Ref.  \cite{kharel2018high}, the acoustic loss in our system is likely dominated by diffraction or a tilt of the crystal relative to the optical axis. Both of these effects can be viewed as a coupling of each resonator mode, whose transverse profile is determined by the optical cavity field, to the transverse modes of the entire crystal. Therefore, these extrinsic, geometric loss mechanisms are most accurately described using the input-output formalism for a resonator coupled to an external bath of modes \cite{walls2007quantum,clerk2010introduction}. In our case, all of the acoustic modes $b_m$ are coupled to a common set of bath modes $c_q$ at frequency $\omega_q$ with rates $f_{m,q}$, giving rise to the coupling Hamiltonian
\begin{align}
    H_c &= -i\hbar\sum_q \sum_m (f_{m,q}b_m^{\dagger} c_q-H.c.)
\end{align}
The coupled equations of motion (EOM) for the bath and resonator modes are then
\begin{align}
    \dot{c}_q &= -i\omega_q c_q+\sum_m f_{m,q}^*b_m\\
    \dot{b}_m &= \frac{i}{\hbar}[H, b_m]+\sum_q f_{m,q}c_q
\end{align}

Following the analysis of Ref.  \cite{clerk2010introduction}, we solve for the EOM of the resonator modes and obtain

\begin{align}
    \dot{b}_m &= \frac{i}{\hbar}[H, b_m]-\frac{\Gamma_{m,e}}{2}b_m-\sum_{n\neq m}\frac{\Gamma_{nm,e}}{2}b_n
\label{eom1}
\end{align}

Here we have made the usual Markov approximation and assumed that the coupling is independent of frequency so that $f_{m,q} = f_m$, which lets us introduce a decay rate $\Gamma_{m,e} = 2\pi\left|f_m\right|^2\rho$, where $\rho = \sum_q\delta(\omega-\omega_q)$ is a constant density of bath states. Furthermore, since the resonator modes we are interested in are different longitudinal modes with the same transverse profile, it is fair to assume that all $f_m$ are approximately equal. For now, we maintain a bit more generality and assume that they all have the same phase, which allows us to define $\Gamma_{nm} = \sqrt{\Gamma_{m,e}\Gamma_{n,e}}$. We emphasize, however, that the sign of the last term in Equation \ref{eom1} is in general determined by the relative phase of the bath coupling for different resonator modes.

In addition to the extrinsic dissipation we have considered so far, acoustic resonators also have intrinsic sources of dissipation such as scattering and absorption. Therefore, to be complete, we introduce an intrinsic decay rate $\Gamma_{m,i}$, so that the total dissipation rate for each acoustic made is $\Gamma_m = \Gamma_{m,e} + \Gamma_{m,i}$, corresponding to the linewidth extracted from OMIT spectra in the limit of weak optomechanical coupling. The final equations of motion for the optical and acoustic modes are then:

\begin{align}
    \dot{a}_m &= \frac{i}{\hbar}[H, a]-\frac{\kappa}{2}a\\
    &=-i\Delta a-i\sum_m g_m b_m-\frac{\kappa}{2}a\\
    \dot{b}_m &= \frac{i}{\hbar}[H, b_m]-\frac{\Gamma_m}{2}b_m-\sum_{n\neq m}\frac{\Gamma_{nm}}{2}b_n\\
    &=-i\Omega_m b_m - i g^*_m a -\frac{\Gamma_m}{2}b_m-\sum_{n\neq m}\frac{\Gamma_{nm}}{2}b_n
\end{align}

To show how these equations of motion give rise to the behavior observed in our experiment, we consider the simple case of two acoustic modes coupled to one optical mode. To further simplify the problem, we assume that the optomechanical couplings and dissipation rates for the two acoustic modes are equal, ie $g_1 = g_2 \equiv g$ and $\Gamma_1=\Gamma_2 \equiv \Gamma$. We also assume that $g$ is real. Furthermore, we assume that $\Omega_1-\Delta = -(\Omega_2-\Delta) \equiv \delta$.  

We now describe how to arrive at the approximate eigenstates of the strongly coupled system and their dissipation rates. The first row of Table \ref{basis1} shows the level diagram and the effective Hamiltonian in the basis of the optical and two acoustic modes $(a, b_1, b_2)$. We then define the ``bright" and ``dark" superpositions of acoustic modes as $B = (b_1+b_2)/\sqrt{2}$ and $D = (b_1-b_2)/\sqrt{2}$. As shown the second row of Table \ref{basis1}, only the bright state directly couples to the optical mode, while dark mode is coupled to the bright mode by $\delta$.
We then make one more basis transformation into the hybridized modes $B_+ = (a+B)/\sqrt{2}$ and $B_- = (a-B)/\sqrt{2}$ under the strong interaction $g$, as shown in the third row of Table \ref{basis1}.

It is now clear that the dark state has two decay channels, giving rise to a total decay rate of
\begin{align}
    \kappa_D&=\frac{\kappa \delta^{2}}{2g^2} + (\Gamma - \Gamma_{12}).
\end{align}
The first term comes from the dark state's couplings $B_+$ and $B_-$, each of which has a decay rate of $\kappa/2$ (under the approximation that $\kappa\gg\Gamma$). As we increase $g$, the dark state's optical component vanishes, and what remains is the second term, which is equal to only the intrinsic loss $\Gamma_i$ of the acoustic modes and can be much smaller than the bare acoustic linewidth $\Gamma$.

\begin{table}
  \centering
  \begin{tabular}{  c  c c }
    \begin{minipage}{.3\textwidth}
      \includegraphics[height = 30mm]{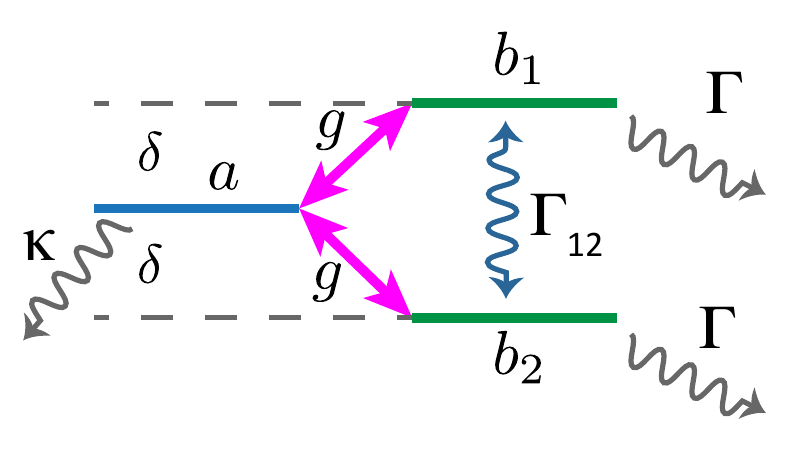}
    \end{minipage}
    &
        $H_1=$
          $\left[\begin{array}{ccc}    
            -i \frac{\kappa}{2} & g & g \\
            g^* & \delta-i\frac{\Gamma}{2} & -i\frac{\Gamma_{12}}{2} \\
            g^* & -i\frac{\Gamma_{12}}{2} & 
            -\delta-i\frac{\Gamma}{2} 
            \end{array}\right]$
            \nonumber
    \\ \hline
    \begin{minipage}{.3\textwidth}
      \includegraphics[height = 30mm]{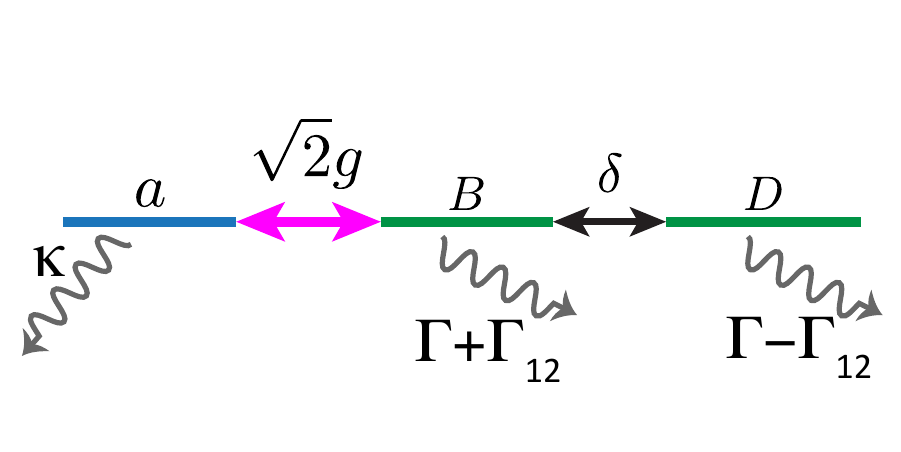}
    \end{minipage}
    &
        $H_2=
         \left[\begin{array}{ccc}    
            -i \frac{\kappa}{2} & \sqrt{2}g & 0 \\
            \sqrt{2}g & -i\frac{\Gamma+\Gamma_{12}}{2} & \delta \\
            0 & \delta & -i\frac{\Gamma-\Gamma_{12}}{2} 
            \end{array}\right]$
            \nonumber
    & 
    \\ \hline
    \begin{minipage}{.3\textwidth}
      \includegraphics[height = 30mm]{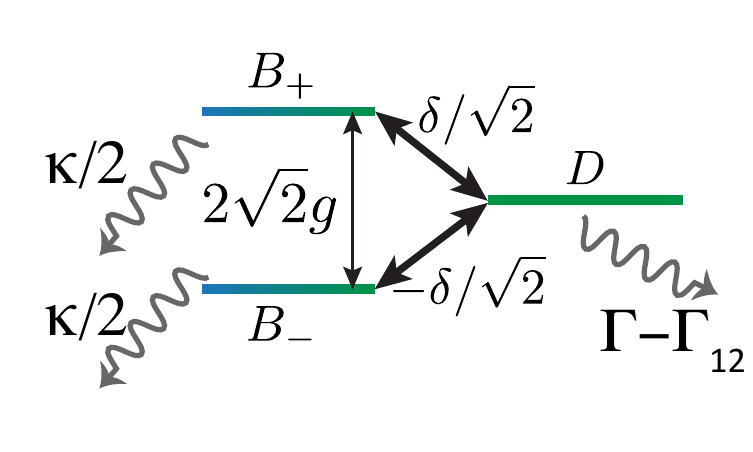}
    \end{minipage}
    &
        $H_3\sim
          \left[\begin{array}{ccc}    
            \sqrt{2}g-i \frac{\kappa}{4} & -i \frac{\kappa}{4} & \frac{\delta}{\sqrt{2}} \\
            -i \frac{\kappa}{4} & -\sqrt{2}g-i \frac{\kappa}{4} & -\frac{\delta}{\sqrt{2}} \\
            \frac{\delta}{\sqrt{2}} & -\frac{\delta}{\sqrt{2}} & -i\frac{\Gamma-\Gamma_{12}}{2} 
            \end{array}\right]$
            \nonumber
    & 
  \end{tabular}
  \caption{Level diagram and effective Hamiltonian in the various bases described in the text. $H_1$ is in the original uncoupled basis of one optical mode and two acoustic modes. By forming symmetric and antisymmetric combinations of the acoustic modes, we arrive at the basis for $H_2$, which already gives the effective dissipation rate for the dark state $D$, as described in the text. The final basis transformation into $H_3$ involves forming the hybridized states of $a$ and $B$ to give $B_{\pm}$, which, together with $D$, would be the actual modes observed in spectroscopy for such a system. For simplicity we have subtracted $\Delta$ from the energies. For $H_3$, we have made the approximation that $\kappa\gg\Gamma, \Gamma_{12}$. }\label{basis1}
\end{table}

Analogously, the case of three acoustic modes strongly coupled to an optical mode leads to the formation of two dark modes, as observed in our experiments. We numerically diagonalize the effective Hamiltonian while adjusting $\Gamma_{12}$ by hand to obtain the theoretical dissipation rate curves in Figure 4d of the main text. We emphasize that our experiment combines three crucial ingredients that are necessary to observe the effect of linewidth narrowing beyond that of the bare acoustic modes. First, the extrinsic loss has to be dominant. Second, the multimode strong coupling to the optical mode effectively eliminates the frequency difference between the acoustic modes so that their extrinsic dissipation channels can interfere. Finally, the bath coupling that leads to this dissipation is approximately equal and in-phase so that this interference is destructive for the dark state that is decoupled from the optical mode in the limit of large $g$. 

Finally, we note such interference in decay channels has also been studied in atoms or resonators coupled to a common reservoir \cite{cardimona1982steady, zhou1997quantum, sundaresan2015beyond}. For example, in the case of atomic excited states radiatively decaying to a common ground state, ultra-narrow spectral lines have been theoretically predicted \cite{zhou1997quantum}. Spectral narrowing has also been observed in multimode microwave resonators coupled to a superconducting qubit \cite{sundaresan2015beyond}, and dissipative coupling terms are also crucial in the analysis of non-reciprocal multimode optomechanical devices that have recently been demonstrated \cite{metelmann2015nonreciprocal, fang2017generalized}.


\end{document}